\documentclass[11pt,a4paper]{article}
\usepackage[T1]{fontenc}
\usepackage{xcolor}
\usepackage{amsmath,amssymb,booktabs,graphicx}
\usepackage[expansion=false]{microtype}
\usepackage{physics}
\usepackage{cite}
\newif\ifjhep
\IfFileExists{jheppub.sty}{\jheptrue}{\jhepfalse}

\ifjhep
  \usepackage{jheppub}
\else
  \usepackage[margin=1.05in]{geometry}
  \usepackage[colorlinks=true,linkcolor=blue,citecolor=blue,urlcolor=blue]{hyperref}
  \newcommand{\JHEPtitle}{}\newcommand{\JHEPauthor}{}
  \newcommand{\JHEPaffil}{}\newcommand{\JHEPabstr}{}\newcommand{\JHEPkeyw}{}
  \renewcommand{\title}[1]{\renewcommand{\JHEPtitle}{#1}}
  \renewcommand{\author}[1]{\renewcommand{\JHEPauthor}{#1}}
  \newcommand{\affiliation}[1]{\renewcommand{\JHEPaffil}{#1}}
  \newcommand{\emailAdd}[1]{\texttt{#1}}
  
  \newcommand{\arxivnumber}[1]{}
  \renewcommand{\abstract}[1]{\renewcommand{\JHEPabstr}{#1}}
  \renewcommand{\maketitle}{%
    \thispagestyle{plain}%
    \begin{center}
      {\LARGE\bfseries\JHEPtitle\par}\vskip 1.4em
      {\large\JHEPauthor\par}\vskip 0.9em
      {\small\itshape\JHEPaffil\par}
    \end{center}
    \vskip 1.2em
    \noindent\textbf{Abstract.}\ \JHEPabstr\par
    \vskip 0.9em
    \noindent{\small\textbf{Keywords:} \JHEPkeyw}\par
    \vskip 1.8em}
  
\fi

\newcommand{\AdS}{\mathrm{AdS}}

\newcommand{\cL}{\mathcal L}

\newcommand{\cP}{\mathcal P}
\newcommand{\cQ}{\mathcal Q}
\newcommand{\al}{\alpha_2}
\newcommand{\be}{\alpha_3}

\newcommand{\CL}{C_{\!\rm L}}

\title{Crossover from Schwarzian dynamics to rotational soft modes in near-extremal Lovelock black holes}

\author{Ankit Anand\footnote{ankitanandp94@gmail.com}$^{a,b}$}
\affiliation{$^{a}$Department of Physics, Indian Institute of Technology Kanpur, Kanpur 208016, India}
\affiliation{$^{b}$Department of Physics, School of Basic Sciences, Indian Institute of Technology Bhubaneswar, Bhubaneswar, Odisha, 752050, India}

\author{Aditya Singh~\footnote{24pr0148@iitism.ac.in}$^{c}$ and}
\affiliation{$^{c}$Department of Physics, Indian Institute of Technology (Indian School of Mines) Dhanbad, Jharkhand 826004, India}

\author{Kimet Jusufi~\footnote{kimet.jusufi@unite.edu.mk}$^{d}$}
\affiliation{$^{d}$Physics Department, State University of Tetovo, Ilinden Street nn, 1200, Tetovo, North Macedonia}

\abstract{Near-extremal black holes are controlled by a small set of soft modes, and counting these modes fixes the coefficient of $\log T$ in the entropy. We ask whether genuine six-derivative interactions change that count, working in seven-dimensional third-order Lovelock gravity, the simplest setting where the cubic Lovelock term is dynamical. The multiplicities stay the same, but the naive count fails. At fixed charge the Maxwell mode is frozen out, and the fifteen rotational modes of the $S^5$ act as a quantum rotor that only switches on above the scale $K_J^{-1}$. As a result, the coefficient of $\log T$ runs from $3/2$ at the lowest temperatures to $9$ above that scale, and never reaches the naive Gaussian value $19/2$. We compute the moment of inertia $K_J$ from the charged slow-rotation solution and find that the Lovelock couplings move the crossover without changing the multiplicities. Only large AdS black holes have a wide window in which the full rotational contribution competes with the classical entropy. Adding a probe charged scalar, we derive an analytic near-horizon Schwinger threshold in the cubic coupling. Its large-AdS limit tells us when the broad rotational crossover window can coexist with the absence of a charged-scalar Breitenlohner--Freedman (BF) bound violation.}

\begin{document}
\maketitle


\section{Introduction}

Black hole thermodynamics is a standard way into questions about quantum gravity. The Bekenstein–Hawking~\cite{Hawking:1971vc, Hawking:1975vcx, Bekenstein:1973ur, Bekenstein:1974ax} area law ties entropy to horizon geometry; quantum and higher-curvature effects show how that relation changes beyond Einstein gravity. In Lovelock theories, the classical entropy is given by the Wald formula~\cite{Wald:1993nt, Wald:1999vt}, which depends on both the higher-curvature couplings and the horizon geometry. Quantum corrections to this entropy are therefore a useful probe of how the microscopic degrees of freedom of a black hole interact with gravitational dynamics. Logarithmic corrections sit between classical gravitational thermodynamics and the ultraviolet completion of gravity. At one loop, the correction is set by the determinant of the quadratic fluctuation operator around the black hole saddle. Such corrections have been studied extensively for several classes of black holes, but in higher-curvature theories one has to account for the modified gravitational dynamics and the accompanying fluctuation spectrum. Extremal geometries are easier to handle because the near-horizon region develops an $\AdS_2$ factor~\cite{Bardeen:1999px, Sen:2007qy}. The quantum entropy function and related heat-kernel methods show that the coefficient of the logarithm can be extracted from infrared data, as long as the zero modes are separated from the ordinary nonzero spectrum and their measure is treated correctly~\cite{Gibbons:1976ue, Banerjee:2010qc, Sen_2012, Sen_20122, Sen:2008vm, Sen:2012dw, Bhattacharyya_2012}.

The near-extremal case is slightly different. At exactly zero temperature, some modes are exact zero modes of the extremal near-horizon case. A small temperature lifts these modes. In the infrared, their eigenvalues then vanish linearly in $T$, and their determinants contribute powers of $\log T$. The resulting low-temperature partition function contains a temperature-dependent contribution, including a $\log T$ term whose coefficient depends on how the soft modes are treated. These corrections are distinct from the usual logarithmic corrections tied to the overall horizon scale. A recent explicit calculation in five-dimensional charged Einstein–Gauss–Bonnet (EGB) gravity~\cite{Alvarado:2026kio} shows how this works: the tensor, vector, and Maxwell sectors were treated separately, and the six Killing vectors of $S^3$ produced six vector towers (see also~\cite{PandoZayas:2026uni, Bueno:2026bby}).

This counting has two problems, which we address here. First, treating an isometry tower as Gaussian and zeta-regularizing it only reproduces the high-temperature limit of the corresponding boundary sector. After reduction, that sector is a particle on a group manifold rather than a set of independent oscillators. Second, the contribution of the Maxwell tower depends on whether the charge or the potential is held fixed. In the fixed-charge ensemble, the Fourier projection freezes the $U(1)$ boundary mode~\cite{Iliesiu:2020qvm, Iliesiu:2022kny}. Both effects change the answer, and they are already visible in four-dimensional Reissner--Nordstr\"om, where the fixed-charge coefficient is $3/2$ for $T\lesssim 1/\CL$, with $\CL$ the coefficient of the linear-in-$T$ entropy defined in Sec.~\ref{sec:extremality}. With both effects understood in four dimensions, we turn to seven-dimensional third-order Lovelock gravity. In $D=7$ the cubic Lovelock density is the first higher-order Euler density that is dynamically nontrivial. A generic curvature-cubed action brings in extra Ostrogradsky-type higher-derivative modes, but Lovelock gravity keeps second-order equations of motion and avoids them~\cite{Woodard:2015zca,Kastor:2010gq,Cai:2006pq,Camanho:2009hu,Bueno:2016dol,Takahashi:2011qda,Kastor:2012se,Mazharimousavi:2008ti,Bardoux:2010sq,Giacomini:2018sho,Singha:2025xah}. This makes $D=7$ a good place to test whether genuine six-derivative curvature interactions change the near-extremal infrared logarithm.

Answering this involves two separate questions. The first is how the Lovelock couplings modify the classical saddle. They change the relation between the charge and the extremal radius, the Wald entropy, the second derivative of the metric function at the extremal horizon, and the coefficient of the linear-in-$T$ entropy correction. The second is whether they change the number of infrared zero-mode towers. On a regular branch, the Lovelock terms deform the quadratic fluctuation operator but add no new independent gauge field, scalar, or higher-derivative metric degrees of freedom. The multiplicity of the geometric zero modes is therefore fixed by the symmetries of the near-horizon saddle. What the Lovelock couplings do set is the kinetic normalization of those sectors, and with it the temperature at which each one switches on, as we show in Sec.~\ref{sec:KJ} and Appendix~\ref{app:KK}. The gravitational vector sector is the subtle one. Because the background is static, one might drop the rotational modes, but they do contribute at one loop. These modes are fluctuations whose internal wave functions are Killing vectors on the compact horizon; they do not describe the background's classical angular momentum. The isometry group of $S^5$ is $SO(6)$, which gives fifteen such vector profiles, the seven-dimensional counterpart of the six Killing vector profiles of $S^3$ in the five-dimensional EGB computation~\cite{Alvarado:2026kio}. Even so, they do not contribute $15/2\log T$ at arbitrarily low temperature. After reduction, they combine into an $SO(6)$ boundary sector whose partition function saturates below its own gap.

The rest of the paper is organized as follows. Section~\ref{sec:theory} reviews third-order Lovelock gravity coupled to a Maxwell field, together with its charged static black holes and their thermodynamics, first in general $D$ and then for $D=7$. Section~\ref{sec:extremality} constructs the extremal near-horizon geometry and the near-extremal expansion, and Sec.~\ref{harmonics} collects the harmonic analysis on $S^5$ that we need. The main computation is in Sec.~\ref{sec:soft}, where we identify the soft modes, treat the gauge sectors exactly and derive the logarithmic correction. In Sec.~\ref{sec:schwinger} we add a charged scalar probe and determine its near-horizon Schwinger threshold analytically. Section~\ref{sec:conclusions} contains our conclusions. Appendix~\ref{app:4D} checks the moment of inertia in four dimensions, and Appendix~\ref{app:KK} solves the charged slow-rotation problem to obtain the rotational susceptibility $K_J$ and the Kaluza--Klein gauge coupling.


\section{Charged black holes in third-order Lovelock gravity}
\label{sec:theory}

\subsection{Action and field equations}

General relativity is diffeomorphism-invariant, and its field equations involve at most second derivatives of the metric. Lovelock gravity keeps both properties in higher dimensions. It is built from curvature invariants whose metric variation still gives equations of motion with no derivatives higher than second order, and up to the usual assumptions behind Lovelock's construction, it is the most general local metric theory with this property that can be formed from the Riemann tensor and the metric~\cite{Lovelock:1971yv, Zumino:1985dp, Padmanabhan:2013xyr}. In higher dimensions, this means the extra curvature terms are nontrivial while the field equations stay second order, as in Einstein gravity. We consider third-order Lovelock gravity in seven spacetime dimensions, minimally coupled to a Maxwell field. Including the boundary contribution, the action is
\begin{equation}
I=\frac{1}{16\pi G_7}\int_{\mathcal M_7} d^7x\,\sqrt{-g}\,
\left[
R-2\Lambda+\alpha_2\mathcal L_2
+\alpha_3\mathcal L_3-F_{\mu\nu}F^{\mu\nu}
\right]+I_{\rm bdy} \ ,
\label{eq:actionfull}
\end{equation}
where \(G_7\) is the seven-dimensional Newton constant and \(\alpha_2\) and \(\alpha_3\) are constant couplings that multiply the quadratic and cubic curvature terms. In $D$ dimensions we write the cosmological constant as~\cite{Kastor:2009wy,Dolan:2010ha,Kubiznak:2012wp}
\begin{equation}
\Lambda=-\frac{(D-1)(D-2)}{2 L^2} \ .
\label{eq:Lambda2}
\end{equation} 
The Gauss–Bonnet invariant is
\begin{equation}
\mathcal L_2=
R_{\mu\nu\rho\sigma}R^{\mu\nu\rho\sigma}
-4R_{\mu\nu}R^{\mu\nu}+R^2 \ .
\label{eq:L2}
\end{equation}
The cubic Lovelock invariant is
\begin{eqnarray}
\mathcal L_3 &=&
2R^{\mu\nu\sigma\kappa}
R_{\sigma\kappa\rho\tau}
R^{\rho\tau}{}_{\mu\nu}
+8R^{\mu\nu}{}_{\sigma\rho}
R^{\sigma\kappa}{}_{\nu\tau}
R^{\rho\tau}{}_{\mu\kappa}
+24R^{\mu\nu\sigma\kappa}
R_{\sigma\kappa\nu\rho}R^\rho{}_\mu
+3R R^{\mu\nu\sigma\kappa}R_{\sigma\kappa\mu\nu}
\nonumber\\
&&+24R^{\mu\nu\sigma\kappa}R_{\sigma\mu}R_{\kappa\nu}
+16R^{\mu\nu}R_{\nu\sigma}R^\sigma{}_\mu-12R R^{\mu\nu}R_{\mu\nu}+R^3 \ .
\label{eq:L3}
\end{eqnarray}
Here \(R_{\mu\nu\rho\sigma}\), \(R_{\mu\nu}\) and \(R\) are the Riemann tensor, Ricci tensor and Ricci scalar. The Maxwell field strength is built from the gauge potential \(A_\mu\) as $F_{\mu\nu}=\partial_\mu A_\nu-\partial_\nu A_\mu.$ The boundary term $I_{\rm bdy}$ contains the Lovelock generalization of the Gibbons–Hawking–York term, which is needed for a well-defined variational principle~\cite{Myers:1987yn, Davis:2002gn}, and the counterterms required when the spacetime is asymptotically AdS~\cite{Balasubramanian:1999re, Dehghani:2006dh}. These terms matter for the full Euclidean action and for defining conserved charges. The infrared logarithm we compute is controlled by the near-horizon region, so the details of the finite counterterm scheme affect only the temperature-independent part of the one-loop free energy.

When the spacetime dimension is too low, the $k^{\rm th}$ Lovelock density is topological or drops out of the local equations of motion entirely. The third-order term is dynamically nontrivial only for $D\geq 7$, so we work in dimensions where the cubic interaction contributes to the bulk equations. Varying the action with respect to the metric and the gauge potential gives the gravitational and Maxwell equations
\begin{eqnarray}
G_{\mu\nu}^{(1)}+\Lambda g_{\mu\nu}
+\alpha_2 G_{\mu\nu}^{(2)}
+\alpha_3 G_{\mu\nu}^{(3)}
= T_{\mu\nu} \qquad ;\qquad
\nabla_\nu F^{\mu\nu}=0 \ , 
\label{EMeq}
\end{eqnarray}
where \(G_{\mu\nu}^{(1)}=R_{\mu\nu}-\frac{1}{2}Rg_{\mu\nu}\) is the Einstein tensor and \(G_{\mu\nu}^{(2)}\) and \(G_{\mu\nu}^{(3)}\) come from the quadratic and cubic Lovelock terms. The electromagnetic stress-energy tensor is \(T_{\mu\nu}=2F_{\mu\rho}F_{\nu}{}^{\rho} -\frac{1}{2}g_{\mu\nu}F_{\rho\sigma}F^{\rho\sigma},\) and the two Lovelock tensors are~\cite{Dehghani:2006dh,Dehghani:2008qr}
\begin{eqnarray*}
G_{\mu \nu }^{(2)} &=&2(-R_{\mu \sigma \kappa \tau }R_{\phantom{\kappa \tau
\sigma}{\nu}}^{\kappa \tau \sigma }-2R_{\mu \rho \nu \sigma }R^{\rho \sigma
}-2R_{\mu \sigma }R_{\phantom{\sigma}\nu }^{\sigma }+RR_{\mu \nu })-\frac{1}{%
2}\mathcal{L}_{2}g_{\mu \nu } \\
G_{\mu \nu }^{(3)} &=&-3(4R^{\tau \rho \sigma \kappa }R_{\sigma \kappa
\lambda \rho }R_{\phantom{\lambda }{\nu \tau \mu}}^{\lambda }-8R_{%
\phantom{\tau \rho}{\lambda \sigma}}^{\tau \rho }R_{\phantom{\sigma
\kappa}{\tau \mu}}^{\sigma \kappa }R_{\phantom{\lambda }{\nu \rho \kappa}%
}^{\lambda }+2R_{\nu }^{\phantom{\nu}{\tau \sigma \kappa}}R_{\sigma \kappa
\lambda \rho }R_{\phantom{\lambda \rho}{\tau \mu}}^{\lambda \rho } \\
&&-R^{\tau \rho \sigma \kappa }R_{\sigma \kappa \tau \rho }R_{\nu \mu }+8R_{%
\phantom{\tau}{\nu \sigma \rho}}^{\tau }R_{\phantom{\sigma \kappa}{\tau \mu}%
}^{\sigma \kappa }R_{\phantom{\rho}\kappa }^{\rho }+8R_{\phantom
{\sigma}{\nu \tau \kappa}}^{\sigma }R_{\phantom {\tau \rho}{\sigma \mu}%
}^{\tau \rho }R_{\phantom{\kappa}{\rho}}^{\kappa } \\
&&+4R_{\nu }^{\phantom{\nu}{\tau \sigma \kappa}}R_{\sigma \kappa \mu \rho
}R_{\phantom{\rho}{\tau}}^{\rho }-4R_{\nu }^{\phantom{\nu}{\tau \sigma
\kappa }}R_{\sigma \kappa \tau \rho }R_{\phantom{\rho}{\mu}}^{\rho
}+4R^{\tau \rho \sigma \kappa }R_{\sigma \kappa \tau \mu }R_{\nu \rho
}+2RR_{\nu }^{\phantom{\nu}{\kappa \tau \rho}}R_{\tau \rho \kappa \mu } \\
&&+8R_{\phantom{\tau}{\nu \mu \rho }}^{\tau }R_{\phantom{\rho}{\sigma}%
}^{\rho }R_{\phantom{\sigma}{\tau}}^{\sigma }-8R_{\phantom{\sigma}{\nu \tau
\rho }}^{\sigma }R_{\phantom{\tau}{\sigma}}^{\tau }R_{\mu }^{\rho }-8R_{%
\phantom{\tau }{\sigma \mu}}^{\tau \rho }R_{\phantom{\sigma}{\tau }}^{\sigma
}R_{\nu \rho }-4RR_{\phantom{\tau}{\nu \mu \rho }}^{\tau }R_{\phantom{\rho}%
\tau }^{\rho } \\
&&+4R^{\tau \rho }R_{\rho \tau }R_{\nu \mu }-8R_{\phantom{\tau}{\nu}}^{\tau
}R_{\tau \rho }R_{\phantom{\rho}{\mu}}^{\rho }+4RR_{\nu \rho }R_{%
\phantom{\rho}{\mu }}^{\rho }-R^{2}R_{\nu \mu })-\frac{1}{2}\mathcal{L}%
_{3}g_{\mu \nu }
\end{eqnarray*}

\subsection{Charged static solutions in $D$ dimensions}
\label{sec:solution}

We now briefly review the static charged black holes of the theory.\footnote{We work in units with $G_D=1$; the Newton constant $G_7$ is restored from Sec.~\ref{sec:7D} onward.} For a static, spherically symmetric ansatz the Maxwell equation can be integrated on its own, independently of the higher-curvature sector, and the electric field is fixed by a single integration constant related to the conserved charge. Solving Eq.~\eqref{EMeq} gives $A=-h(r)\,\dd t$ with \(h(r)=q/[(D-3)r^{D-3}],\) where \(q\) is that integration constant. Substituting this into the gravitational equations with the static metric ansatz reduces the Einstein–Lovelock system to a radial equation for \(f(r)\),
\begin{eqnarray}
&&\left[180\binom{D-2}{5}\alpha_3 r(f-1)^2-6\binom{D-2}{3}\alpha_2(f-1)r^3+\frac{D-2}{2}r^5\right]f'+\binom{D-2}{2}r^4(f-1)\nonumber\\
&&-12\binom{D-2}{4}\alpha_2r^2(f-1)^2 +360\binom{D-2}{6}\alpha_3(f-1)^3 +\Lambda r^6 =-q^2r^{10-2D}.
\label{Eqf}
\end{eqnarray}
Here \(f'=df/dr\), and the binomial coefficients carry the dimension dependence of the Lovelock terms. This is the standard Einstein–Lovelock form~\cite{Dehghani:2006dh}. For generic couplings, one radial integration turns it into a cubic algebraic equation for a combination of $f(r)$. We take
\begin{equation}
\alpha _{3}=\frac{\alpha ^{2}}{72(_{\phantom{d}{4}}^{D-3})},\hspace{1cm}%
\alpha _{2}=\frac{\alpha }{(D-3)(D-4)}  \label{con}
\end{equation}
for which the cubic has one real root and a complex-conjugate pair. The real solution of Eq.~(\ref{Eqf}) on this locus is~\cite{Dehghani:2005vh}
\begin{equation}
f(r)=1+{\frac{{r}^{2}}{\alpha }}\left\{ 1-\left( {1+\frac{6\Lambda \alpha }{(D-1)(D-2)}+\frac{3\alpha \tilde m}{(D-2)r^{D-1}}-\frac{6\alpha {q}^{2}}{\left(D-2\right) (D-3)r^{2D-4}}}\right) ^{1/3}\right\}  \label{fsp}
\end{equation}
where $\tilde m$ is the mass parameter. The tilde distinguishes it from the parameter $m$ of Eq.~\eqref{eq:lovelockpoly}; comparing the two gives $\tilde m=(D-2)\,m$, so $\tilde m=5m$ at $D=7$. The solution is asymptotically flat for $\Lambda=0$, asymptotically AdS for $\Lambda<0$ and asymptotically dS for $\Lambda>0$. For these charged solutions the curvature invariants generally diverge at $r=0$, which signals a curvature singularity there.

\subsection{Thermodynamics}
\label{Therm}

The Hawking temperature follows from regularity of the Euclidean section. Continuing $t\rightarrow i\tau$ gives a Euclidean manifold, and avoiding a conical singularity at the horizon requires $\tau\sim\tau+\beta_+$, where $\beta_+=1/T$ and $T=f'(r_+)/4\pi$. On the horizon, $f(r_+)=0$ sets the bracket in Eq.~(\ref{fsp}) equal to $(1+\alpha/r_+^2)^3$, so the cube roots drop out and the temperature depends only on the horizon radius and the charge,
\begin{equation}
T=\frac{(D-2)\,r_+^{2D-10}\left[3(D-3)r_+^4+3(D-5)\alpha r_+^2+(D-7)\alpha^2\right]-6q^2}{12\pi(D-2)\,(r_+^2+\alpha)^2\,r_+^{2D-9}}
-\frac{\Lambda\,r_+^5}{2\pi(D-2)\,(r_+^2+\alpha)^2}\ .
\label{TempH}
\end{equation}
For $\alpha=q=\Lambda=0$ this reduces to the Schwarzschild--Tangherlini value $T=(D-3)/(4\pi r_+)$.

In Einstein gravity the entropy of a stationary black hole is a quarter of the horizon area, in units where the gravitational coupling is suitably normalized, and this holds throughout two-derivative Einstein gravity. Higher-curvature theories add contributions from curvature invariants intrinsic to the horizon~\cite{Zumino:1985dp,Padmanabhan:2013xyr,Jacobson:1993xs}. In Lovelock gravity these corrections are captured by the Jacobson--Myers entropy functional~\cite{Jacobson:1993xs,Wald:1993nt,Iyer:1994ys}, so once the higher-order densities are present the entropy is no longer fixed by the area alone. For a general Lovelock theory, the entropy of a stationary Killing horizon is
\begin{equation}
S=\frac{1}{4} \sum_{k=1}^{[(D-1)/2]}k\alpha _{k}\int d^{D-2}x \sqrt{\tilde{g}} \tilde{\mathcal{L}}_{k-1}  \label{Enta}
\end{equation}
where $\alpha_1=1$ and the integral runs over a spatial cross section of the Killing horizon. Here $\tilde{g}_{ab}$ is the induced metric on that cross section, $\tilde g$ is its determinant, and $\tilde{\mathcal L}_{k}$ is the Lovelock density built from the intrinsic curvature of the induced geometry. This form shows directly how higher-curvature interactions change the entropy even though the gravitational equations stay second order~\cite{Jacobson:1993xs,Iyer:1994ys,Dehghani:2008qr,Padmanabhan:2013xyr}. For third-order Lovelock theory, the Wald formula gives the entropy in terms of intrinsic curvature invariants of the horizon,
\begin{equation}
S=\frac{1}{4}\int d^{D-2}x\,\sqrt{\tilde g}\,
\left[
1+2\alpha_2\tilde R
+3\alpha_3\left(
\tilde R_{\mu\nu\rho\sigma}\tilde R^{\mu\nu\rho\sigma}
-4\tilde R_{\mu\nu}\tilde R^{\mu\nu}
+\tilde R^2
\right)
\right],
\label{Entb}
\end{equation}
where the tilded tensors are the Riemann tensor, Ricci tensor and Ricci scalar of the induced metric \(\tilde g_{ab}\) on a spatial cross-section of the event horizon. The entropy is therefore corrected by both the intrinsic Ricci scalar and the Gauss–Bonnet invariant of the horizon. For a constant-curvature horizon these invariants reduce to powers of the horizon radius, and
\begin{equation}
S=\frac{1}{4}r_{+}^{D-2}\Sigma _{D-2}\left( 1+\frac{4\alpha _{2}(_{\phantom{D}{2}}^{D-2})}{r_{+}^{2}}+\frac{72(_{\phantom{D}{4}}^{D-2})\alpha_{3}}{r_{+}^{4}}\right)  \label{Ent1}
\end{equation}
where $\Sigma_{D-2}$ is the area of a unit $(D-2)$-dimensional sphere. Switching off the higher-curvature couplings recovers the Einstein area law.

The conserved electric charge is the electric flux through a large spatial surface. With the normalization of the Maxwell term in the action,
\begin{equation}
\cQ =\frac{\Sigma _{D-2}}{4\pi }q  \label{Ch}
\end{equation}
The electrostatic potential, measured relative to the horizon, is
\begin{equation}
\Phi =A_{\mu }\chi ^{\mu }\left| _{r\rightarrow \infty }-A_{\mu }\chi ^{\mu}\right| _{r=r_{+}},  \label{Pot1}
\end{equation}
where $\chi=\partial/\partial t$ is the horizon-generating Killing vector. For the static configuration, with $A=-h(r)\,\dd t$, this gives
\begin{equation}
\Phi =\frac{q}{(D-3)r_{+}^{D-3}} \ .
\label{Pot2}
\end{equation}
The mass follows from the asymptotic behavior of the metric. In the normalization above,
\begin{equation}
M=\frac{\Sigma _{D-2}}{16\pi }\tilde m; \quad \tilde m= (D-2) r_+^{D-3}+\frac{2 q^2}{(D-3)r_+^{D-3}}+\frac{1}{3} \alpha (D-2) r_+^{D-7} \left(\alpha +3 r_+^2\right) -\frac{2 \Lambda  r_+^{D-1}}{(D-1)}\label{Mass1}
\end{equation}
These quantities can be checked against the first law. From the entropy, charge and mass in Eqs.~(\ref{Ent1}), (\ref{Ch}) and (\ref{Mass1}), it is straightforward to compute $\partial S/\partial r_+$, $\partial \cQ/\partial r_+$ and $\partial M/\partial r_+$. Holding the Lovelock couplings and the cosmological constant fixed, one finds
\begin{equation}
T=\left( \frac{\partial M}{\partial S}\right) _{\cQ},\ \ \ \ \Phi =\left(\frac{\partial M}{\partial \cQ}\right) _{S}  \label{Dsmar}
\end{equation}
which reproduce the temperature and potential of Eqs.~(\ref{TempH}) and (\ref{Pot2}). The thermodynamic variables are therefore consistent and satisfy
\begin{equation}
dM=TdS+\Phi d\cQ
\label{1stlaw}
\end{equation}
This relation fixes the thermodynamic normalization used in the near-horizon analysis. Note that not all of the expressions above share the same range of validity. The solution~\eqref{fsp}, the temperature~\eqref{TempH} and the mass~\eqref{Mass1} hold only on the special coupling locus Eq.~\eqref{con}, while the entropy, Eqs.~\eqref{Enta}--\eqref{Ent1}, the charge, Eq.~\eqref{Ch}, and the potential, Eq.~\eqref{Pot2}, hold for arbitrary couplings. From now on, we drop this restriction and assume $\alpha_2$ and $\alpha_3$ are independent.
\subsection{The seven-dimensional solution}
\label{sec:7D}

We now set $D=7$. In seven dimensions, the third-order Lovelock density is the highest-order Lovelock term that contributes to the local equations of motion and $D=7$ is the lowest dimension in which the cubic term is dynamical. We take a static configuration whose horizon is a five-dimensional constant-curvature manifold, with metric~\cite{Cai:2006pq, Kastor:2010gq, Garraffo:2008hu, Padmanabhan:2013xyr}
\begin{equation}
ds^2=-f(r)\dd t^2+\frac{\dd r^2}{f(r)}+r^2d\Sigma_{5,k}^2 \ ,
\label{eq:metricstatic}
\end{equation}
where $d\Sigma_{5,k}^2$ is the line element of a five-dimensional space of constant sectional curvature $k$, whose volume we denote by $\Sigma_{5,k}$. For the spherical case $k=1$ this space is the unit $S^5$, with $\Sigma_{5,1}=\Omega_5=2\pi^{3}/\Gamma(3)=\pi^3$, which is $\Sigma_{D-2}$ of Sec.~\ref{Therm} at $D=7$. The Maxwell equations integrate immediately in the static ansatz, yielding the radial electric field $F_{rt}=q/r^5$. We introduce the charge parameter $Q$ through $q=\sqrt{10}\,Q$, which is the statement $2q^2/((D-2)(D-3))=Q^2$ at $D=7$. The metric function then satisfies the polynomial
\begin{equation}
\frac{k-f(r)}{r^2}
+a\left(\frac{k-f(r)}{r^2}\right)^2
+b\left(\frac{k-f(r)}{r^2}\right)^3
=-\frac{1}{L^2}+\frac{m}{r^6}-\frac{Q^2}{r^{10}} \ ,
\label{eq:lovelockpoly}
\end{equation}
where $a=12\al$ and $b=24\be$. These are the dimensionally dressed couplings at $D=7$, namely $a=(D-3)(D-4)\al$ and $b=(D-3)(D-4)(D-5)(D-6)\be$. The length $L$ enters through Eq.~\eqref{eq:Lambda2}, but it is a bare parameter, not the asymptotic AdS radius. For $\al,\be\neq0$ the effective AdS radius $L_{\rm eff}$ is the Einstein-branch root of $-L_{\rm eff}^{-2}+aL_{\rm eff}^{-4}-bL_{\rm eff}^{-6}=-L^{-2}$, and it reduces to $L$ only when $a=b=0$. All formulas below are written in terms of Eq.~\eqref{eq:lovelockpoly}. To avoid a clash of notation, we reserve $\cL_2$ for the Gauss–Bonnet density and $L_2$ for the $\AdS_2$ radius introduced in Eq.~\eqref{eq:L2sphere}; the two are unrelated. With this normalization, a maximally symmetric space of sectional curvature $K$ in $D$ dimensions has $\cL_p=K^p D!/(D-2p)!$, a check we use repeatedly below.

The parameter $Q$ is proportional to the conserved charge, $\cQ=\sqrt{10}\,\Sigma_{5,k}\, Q/(4\pi G_7)$, so fixing $Q$ is the same as fixing $\cQ$, and we use the two interchangeably for the fixed-charge ensemble. The Einstein branch is selected by requiring continuity at $a=b=0$~\cite{Boulware:1985wk, Wheeler:1985nh}. The other algebraic branches are not continuously connected to Einstein gravity and can have vanishing or negative effective graviton kinetic terms. At $r=r_+$, where $f(r_+)=0$, Eq.~\eqref{eq:lovelockpoly} gives
\begin{equation}
m=r_+^4\left(k+\frac{ak^2}{r_+^2}+\frac{bk^3}{r_+^4} +\frac{r_+^2}{L^2}+\frac{Q^2}{r_+^8} \right) =kr_+^4+ak^2r_+^2+bk^3+\frac{r_+^6}{L^2}+\frac{Q^2}{r_+^4} \ .
\label{eq:mass}
\end{equation}
Differentiating Eq.~\eqref{eq:lovelockpoly} and evaluating the result at the horizon gives the temperature $T=f'(r_+)/4\pi$,
\begin{equation}
T=\frac{6r_+^6/L^2+4kr_+^4+2ak^2r_+^2-4Q^2/r_+^4}{4\pi r_+\left(r_+^4+2akr_+^2+3bk^2\right)}=\frac{6r_+^6/L^2+4kr_+^4+24\al k^2r_+^2-4Q^2/r_+^4}{4\pi r_+\left(r_+^4+24\al kr_+^2+72\be k^2\right)} \ .
\label{eq:Tfull}
\end{equation}
The Wald entropy follows from the Lovelock entropy formula~\cite{Wald:1993nt, Jacobson:1993xs}. For $D=7$,
\begin{equation}
S_{\rm W}(r_+)=\frac{\Sigma_{5,k}}{4G_7}
\left(r_+^5+40\al kr_+^3+360\be k^2r_+\right) \ ,
\label{eq:Wald}
\end{equation}
and for $k=1$,
\begin{equation}
S_{\rm W}(r_+)=\frac{\pi^3}{4G_7}
\left(r_+^5+40\al r_+^3+360\be r_+\right) \ .
\label{eq:Waldsphere}
\end{equation}
The coefficients follow from
\begin{equation*}
    S=\frac{1}{4G_7}\int \sqrt{h} \;\sum_p p\,c_p\,\cL_{p-1}(h) \qquad \text{with} \qquad\cL_p=\frac{(D-2)!}{(D-2-2p)!}\left(\frac{k}{r_+^2}\right)^p \ ,
\end{equation*}
where $c_1=1$, $c_2=\al$ and $c_3=\be$, which gives $40\al$ and $360\be$. At fixed Lovelock couplings the first law is
\begin{equation}
\dd M=T\dd S_{\rm W}+\Phi\dd \cQ,
\label{eq:firstlaw}
\end{equation}
with $M=5\,\Sigma_{5,k}\,m/(16\pi G_7)$, $\cQ=\sqrt{10}\,\Sigma_{5,k}\,Q/(4\pi G_7)$ and $\Phi=q/(4r_+^4)$.


\section{Extremality and the exact near-horizon solution}
\label{sec:extremality}

In this section we take the extremal limit of the seven-dimensional solution, find its exact near-horizon geometry, and then expand around it at small temperature. We begin with the extremality conditions. An extremal horizon at $r=r_0$ is a double zero of the metric function, so that $f(r_0)=0$ and $f'(r_0)=0$. The second condition is equivalent to $T(r_0)=0$, and Eq.~\eqref{eq:Tfull} then gives
\begin{equation}
Q_0^2=kr_0^8+\frac{a}{2}k^2r_0^6+\frac{3}{2}\frac{r_0^{10}}{L^2} = kr_0^8+6\al k^2r_0^6+\frac{3r_0^{10}}{2L^2} \ .
\label{eq:Q0}
\end{equation}
The cubic coupling does not occur in this relation in seven dimensions. It is nevertheless present in the extremal mass and entropy, and it enters the near-horizon radius through the denominator of $f''(r_0)$. Substituting Eq.~\eqref{eq:Q0} into Eq.~\eqref{eq:mass}, we obtain
\begin{equation}
m_0=2kr_0^4+\frac32ak^2r_0^2+bk^3+\frac52\frac{r_0^6}{L^2} = 2kr_0^4+18\al k^2r_0^2+24\be k^3+\frac52\frac{r_0^6}{L^2} \ .
\label{eq:m0ab}
\end{equation}
Expanding the metric function about the double root gives
\begin{equation}
f(r)=\frac12f''(r_0)(r-r_0)^2+\mathcal{O}((r-r_0)^3) \ .
\label{eq:fexpand}
\end{equation}
A direct differentiation of the Lovelock polynomial gives
\begin{equation}
f''(r_0)=\frac{4\left(15r_0^4+8kL^2r_0^2+3ak^2L^2\right)}{L^2\left(r_0^4+2akr_0^2+3bk^2\right)} \ .
\label{eq:fppab}
\end{equation}
We note for later use that $b$ cancels identically in the numerator of Eq.~\eqref{eq:fppab}. Differentiating Eq.~\eqref{eq:lovelockpoly} twice at the double root gives $-g_1f''/r_0^2=-40Q_0^2/r_0^{12}+8k/r_0^4+8ak^2/r_0^6$ with $g_1=1+2ak/r_0^2+3bk^2/r_0^4$, so the cubic coupling survives only through $g_1$. The $\AdS_2$ radius then follows from $L_2^2=2/f''(r_0)$, and for the spherical saddle it takes the form
\begin{equation}
L_2^2=\frac{L^2\left(r_0^4+24\al r_0^2+72\be\right)}{2\left(15r_0^4+8L^2r_0^2+36\al L^2\right)}\; = \frac{L^2\Delta}{2W},
\label{eq:L2sphere}
\end{equation}
where throughout we abbreviate
\begin{equation}
\Delta= r_0^4+24\al r_0^2+72\be,
\qquad
W= 15r_0^4+8L^2r_0^2+36\al L^2 \ .
\label{eq:DeltaW}
\end{equation}
As a consistency check, in the asymptotically flat Einstein limit, $\alpha_2=\alpha_3=0$ and $L\to\infty$, one finds $L_2^2\to r_0^2/16=r_0^2/(D-3)^2$, the known extremal Reissner--Nordstr\"om value in $D$ dimensions. The Lorentzian near-horizon limit~\cite{Bardeen:1999px, Sen:2007qy} is obtained by defining
\begin{equation}
r=r_0+\epsilon\rho, \qquad t=\frac{L_2^2}{\epsilon}\tau, \qquad \epsilon\to 0 \ ,
\label{eq:nhscaling}
\end{equation}
and keeping $(\rho,\tau)$ fixed. The resulting geometry is
\begin{equation}
 ds^2_{\rm NH}=L_2^2\left(-\rho^2\dd\tau^2+\frac{\dd\rho^2}{\rho^2}\right)+r_0^2d\Omega_5^2 \ .
\label{eq:NH}
\end{equation}
The field strength becomes constant on the $\AdS_2$ factor as
\begin{equation}
F_{\rm NH}=e\,\dd\rho\wedge\dd\tau,
\qquad
 e=\frac{q_0L_2^2}{r_0^5}
=\frac{\sqrt{10}Q_0L_2^2}{r_0^5}.
\label{eq:NHF}
\end{equation}
The additive constant in $A_\tau$ is gauge dependent; the electric field $e$ is not. The Euclidean continuation $\tau\rightarrow-i\tau_E$ gives
\begin{equation}
ds^2_{\rm NH,E}=L_2^2\left(\rho^2\dd\tau_E^2+\frac{\dd\rho^2}{\rho^2}\right)+r_0^2d\Omega_5^2.
\label{eq:NHE}
\end{equation}
For the zero-temperature throat $\tau_E$ is noncompact. The near-extremal black hole replaces the Poincaré throat by a finite-temperature $\AdS_2$ black hole~\cite{Rakic:2023vhv,Heydeman:2020hhw,Ferrara:1995ih,Banerjee:2023gll,Banerjee:2023quv,Banerjee:2026pad}.

To describe the near-extremal geometry and thermodynamics, we expand the horizon radius about $r_0$ as
\begin{equation}
r_+=r_0+\delta r \ , \qquad 0<\delta r\ll r_0 \ .
\label{eq:deltar}
\end{equation}
At fixed charge $Q=Q_0$, the temperature is obtained by expanding Eq.~\eqref{eq:Tfull}. Since the numerator vanishes at $r_0$ and the denominator is finite on a regular branch,
\begin{equation}
T=\frac{f''(r_0)}{4\pi}\delta r+O(\delta r^2) =\frac{\delta r}{2\pi L_2^2}+O(\delta r^2) \Longrightarrow \delta r=2\pi L_2^2T+O(T^2) \ .
\label{eq:deltarT}
\end{equation}
Now, expanding the Wald entropy,
\begin{align}
S_{\rm W}(r_+)
&=S_0+\left.\frac{\partial S_{\rm W}}{\partial r_+}\right|_{r_0}\delta r+O(\delta r^2)=S_0+\frac{5\pi^3\Delta}{4G_7}\,\delta r+O(\delta r^2) \ .
\label{eq:Sexpand1}
\end{align}
Using Eq.~\eqref{eq:deltarT}, this can be written as
\begin{equation}
S_{\rm W}(T)=S_0+4\pi^2\CL T+O(T^2) \quad \text{where}\quad \CL= \frac{5\pi^2\Delta L_2^2}{8G_7}\;=\;\frac{L_2^2}{2\pi}\frac{\partial S_{\rm W}}{\partial r_+}\bigg|_{r_0} \ .
\label{eq:CL}
\end{equation}
This $\CL$ is the coefficient which controls the classical near-extremal energy above extremality and the Schwarzian sector. The heat capacity is $C_Q=T(\partial S/\partial T)_Q=4\pi^2\CL T+O(T^2)$, and integrating it gives the energy above extremality, $E-E_0=2\pi^2\CL T^2+O(T^3)$. The canonical near-extremal partition function has the classical form~\cite{Preskill:1991tb,Iliesiu:2020qvm},
\begin{equation}
\log Z_{\rm cl} = S_0+2\pi^2\CL T+O(T^2) \ ,
\label{eq:Zcl}
\end{equation}
up to the conventional subtraction of the extremal energy in the Boltzmann factor. More precisely, the universal Schwarzian effective action~\cite{Maldacena:2016upp,Stanford:2017thb} produces
\begin{equation}
Z_{\rm Sch}\propto \left(\CL T\right)^{3/2} \exp\left(2\pi^2\CL T\right) \ ,
\label{eq:SchZ}
\end{equation}
where the power $3/2$ is the one-loop determinant of the boundary reparametrization modes. The overall proportionality constant is not fixed by the infrared mode counting. Equations~\eqref{eq:Zcl} and~\eqref{eq:SchZ} are not independent: the prefactor $(\CL T)^{3/2}$ in Eq.~\eqref{eq:SchZ} is the same factor produced by the Schwarzian tower, Eq.~\eqref{eq:ZSch}, and $Z_{\rm cl}$ in Eq.~\eqref{eq:Zcl} denotes only the exponential. Matching Eq.~\eqref{eq:SchZ} to the standard Schwarzian normalization $Z=(\Phi_r/\beta)^{3/2}e^{2\pi^2\Phi_r/\beta}$ identifies $\Phi_r=\CL$. For completeness, a convenient finite-temperature $\AdS_2$ black-hole form is~\cite{Alvarado:2026kio,PandoZayas:2026uni,Rakic:2023vhv,Banerjee:2026pad,Iliesiu:2022onk},
\begin{equation}
ds^2_{2,T}=L_2^2\left[(\rho^2-\rho_h^2)\dd\tau^2+\frac{\dd\rho^2}{\rho^2-\rho_h^2}\right],
\label{eq:AdS2BH}
\end{equation}
with Euclidean period $\beta=2\pi/\rho_h$ and $\rho_h=2\pi T$. This corresponds to the choice $\epsilon=L_2^2$ in Eq.~\eqref{eq:nhscaling}: then $\rho_h=\delta r/\epsilon=2\pi T$, and the Euclidean period of $\tau$ equals that of the original time $t$, namely $1/T$. Thus
\begin{equation}
ds^2_{\rm NH,T}=L_2^2\left[(\rho^2-(2\pi T)^2)\dd\tau_E^2 +\frac{\dd\rho^2}{\rho^2-(2\pi T)^2}\right] +r_0^2d\Omega_5^2+O(T) \ ,
\label{eq:NHnearT}
\end{equation}
where $\tau_E\sim\tau_E+1/T$. Other choices of $\epsilon$ rescale $\rho$ and $\tau$ and do not modify the zero-mode counting~\cite{Singh:2023hit,Castro:2018hsc,Bardeen:1999px,Larsen:2014bqa}.

\section{Harmonic decomposition on the internal $S^5$}\label{harmonics}

The near-horizon geometry contains a compact five-sphere. The Kaluza–Klein decomposition of the metric and ghost fluctuations is therefore most conveniently organized in terms of scalar, transverse-vector, and transverse-traceless tensor harmonics on $S^5$. We use a unit-radius five-sphere throughout this section and denote its metric and covariant derivative by $\hat g_{ij}$ and $\hat D_i$, respectively. The unit five-sphere may be embedded in $\mathbb{R}^6$ as $X^A X_A=1$, with $A=1,\ldots,6$ and induced metric
\begin{equation}
\hat g_{ij} = \partial_i X^A\,\partial_j X_A .
\end{equation}
Its curvature tensors are $\hat R_{ij}=4\hat g_{ij}$,\, $\hat R=20$ and $\Omega_5=\pi^3$. The embedding coordinates satisfy $\hat D_i\hat D_jX^A = -\hat g_{ij}X^A$,\, $\hat D^2X^A=-5X^A$ and $\hat D_iX^A\hat D^iX^B = \delta^{AB}-X^AX^B $. A complete set of scalar harmonics can be constructed by restricting homogeneous harmonic polynomials in $\mathbb{R}^6$ to the unit sphere~\cite{Rubin:1984tc,Higuchi:1986wu},
\begin{equation}
Y^{(l,\mathcal A)}(X)=C^{(\mathcal A)}_{A_1\cdots A_l}
X^{A_1}\cdots X^{A_l} \quad\text{where}\quad C^{(\mathcal A)}_{A_1\cdots A_l} = C^{(\mathcal A)}_{(A_1\cdots A_l)} \quad \delta^{A_1A_2} C^{(\mathcal A)}_{A_1A_2A_3\cdots A_l} = 0 \ .
\end{equation}
They obey
\begin{equation}
-\hat D^2Y^{(l,\mathcal A)} = l(l+4)Y^{(l,\mathcal A)} \qquad l=0,1,2,\ldots \ .
\end{equation}
The degeneracy can be computed as
\begin{equation}
d_l^{(0)} = \binom{l+5}{l} - \binom{l+3}{l-2} = \frac{(l+1)(l+2)^2(l+3)}{12} \ ,
\end{equation}
where the second binomial coefficient is understood to vanish for $l=0,1$. The unique scalar Laplacian zero mode is the constant harmonic \(Y^{(0)} = 1/\sqrt{\pi^3} \quad ; \quad -\hat D^2Y^{(0)}=0  \). The six $l=1$ scalar harmonics ($A=1,\ldots,6$) are
\begin{equation}
Y^{(1),A}=X^A \qquad \text{and}\qquad -\hat D^2X^A=5X^A \ .
\end{equation}
They are consequently not scalar Laplacian zero modes. Their normalization follows from
\begin{equation}
\int_{S^5}d\Omega_5\,X^AX^B = \frac{\pi^3}{6}\delta^{AB} \ ,
\end{equation}
so an orthonormal basis is $\mathcal Y^A = \sqrt{6/\pi^3}\,X^A$.

The vector harmonics split into scalar-derived and transverse sectors.  The longitudinal vector harmonics are
\begin{equation}
Y_i^{L,(l,\mathcal A)} = \hat D_iY^{(l,\mathcal A)} \ .
\end{equation}
Their divergence is
\begin{equation}
\hat D^iY_i^{L,(l,\mathcal A)} = -l(l+4)Y^{(l,\mathcal A)} \ .
\end{equation}
Using $\left[\hat D^2,\hat D_i\right]Y = \hat R_i{}^j\hat D_jY = 4\hat D_iY$, one obtains the rough-Laplacian eigenvalue
\begin{equation}
-\hat D^2Y_i^{L,(l,\mathcal A)} = \left[l(l+4)-4\right] Y_i^{L,(l,\mathcal A)} \ .
\end{equation}
The corresponding eigenvalue of the Hodge--de Rham operator $\Box_{\rm H}=\mathrm d\delta+\delta\mathrm d$ is instead
\begin{equation}
\Box_{\rm H}Y_i^{L,(l,\mathcal A)} = l(l+4)Y_i^{L,(l,\mathcal A)} .
\end{equation}
This distinction is important because the rough Laplacian and the Hodge--de Rham operator differ on one-forms by the Ricci curvature. The transverse vector harmonics satisfy
\begin{equation}
\hat D^iY_i^{T,(l,\mathcal A)}=0 \quad \text{and}\quad -\hat D^2Y_i^{T,(l,\mathcal A)}
=
\left[l(l+4)-1\right]
Y_i^{T,(l,\mathcal A)} ,
\qquad
\text{for}\quad l=1,2,\ldots .
\end{equation}
Their degeneracy is
\begin{equation}
d_l^{(1)}=\frac{l(l+4)(2l+4)(l+2)}{6} = \frac{l(l+2)^2(l+4)}{3} \ .
\end{equation}
The $l=1$ transverse vector harmonics are the fifteen Killing vectors of $S^5$.  In terms of the embedding coordinates, we can write
\begin{equation}\label{Eq:Kiling_vectors}
K_i^{AB} = X^A\hat D_iX^B-X^B\hat D_iX^A \ , \qquad A<B \ .
\end{equation}
There are $\binom{6}{2}=15$ independent vectors, generating the Lie algebra $\mathfrak{so}(6)\cong\mathfrak{su}(4)$. Globally, $\mathrm{Spin}(6)\cong SU(4)$ and $SO(6)\cong SU(4)/\mathbb{Z}_2$. We use the connected geometric rotation group $SO(6)$ for the bosonic rotor sector. On direct differentiation, we get
\begin{equation}
\hat D_iK_j^{AB} = \hat D_iX^A\hat D_jX^B -\hat D_iX^B\hat D_jX^A \ ,
\end{equation}
and hence
\begin{equation}
\hat D_{(i}K_{j)}^{AB}=0 \ , \qquad \hat D^iK_i^{AB}=0 \ .
\end{equation}
The rough-Laplacian eigenvalue is $-\hat D^2K_i^{AB}=4K_i^{AB}$. Their normalization is
\begin{equation}\label{eq:Knorm}
\int_{S^5}d\Omega_5\, K_i^{AB}K^{i,CD} = \frac{\pi^3}{3}\left(\delta^{AC}\delta^{BD}-\delta^{AD}\delta^{BC}\right) \ .
\end{equation}
Thus, one may define normalized Killing vectors by
\begin{equation}
\mathcal K_i^{AB} = \sqrt{\frac{3}{\pi^3}}\,K_i^{AB} \ .
\end{equation}
The Killing vectors must not be confused with zero modes of the rough vector Laplacian. Their metric variation vanishes,
\begin{equation}
\delta_K\hat g_{ij} = 2\hat D_{(i}K_{j)} = 0 \ ,
\end{equation}
so they represent infinitesimal isometries.  Their treatment in a gravitational path integral depends on the gauge-fixing prescription and on whether the corresponding gauge-group volume and Faddeev–Popov zero modes are divided out~\cite{Sen:2012dw,Dabholkar:2010rm,Dabholkar:2012zz,Iliesiu:2022onk}.
The genuine tensor sector consists of symmetric transverse-traceless harmonics~\cite{Rubin:1984tc,Higuchi:1986wu},
\begin{equation}
Y_{ij}^{TT,(l,\mathcal A)}  = Y_{ji}^{TT,(l,\mathcal A)} \ ,
\end{equation}
satisfying $\hat D^iY_{ij}^{TT,(l,\mathcal A)}=0$ and $\hat g^{ij}Y_{ij}^{TT,(l,\mathcal A)}=0$.  There are no nontrivial TT tensor harmonics for $l=0,1$, and therefore $l$ takes $2,3,\cdots$. Their rough-Laplacian eigenvalues are
\begin{equation}
-\hat D^2Y_{ij}^{TT,(l,\mathcal A)}
=
\left[l(l+4)-2\right]
Y_{ij}^{TT,(l,\mathcal A)} .
\end{equation}
The degeneracy is
\begin{equation}
d_l^{(2)} = \frac{3}{4}(l-1)(l+2)^2(l+5) \ ,
\qquad l\geq2 \ .
\end{equation}
It is useful to distinguish the genuine TT sector from tensor structures constructed from scalar harmonics.  For an $l=2$ scalar harmonic satisfying $-\hat D^2Y^{(2)}=12Y^{(2)}$,  the traceless scalar-derived tensor
\begin{equation}
H_{ij} = \hat D_i\hat D_jY^{(2)} + \frac{12}{5}\hat g_{ij}Y^{(2)}
\end{equation}
obeys $\hat g^{ij}H_{ij}=0$,  but
\begin{equation}
\hat D^iH_{ij}=-\frac{28}{5}\hat D_jY^{(2)}\neq0  \ .
\end{equation}
The distinction between the spherical harmonic spectrum and the spectrum of the complete quadratic operator is essential.  In particular,
\begin{equation}
-\hat D^2Y^{(0)}=0 \ , \qquad -\hat D^2X^A=5X^A \ , \qquad -\hat D^2K_i^{AB}=4K_i^{AB} \ ,
\end{equation}
so the constant scalar, the six $l=1$ scalar harmonics, and the fifteen Killing vectors have different mathematical meanings~\cite{Rubin:1984tc,Higuchi:1986wu,Kodama:2003jz}. Only after the complete Lovelock Hessian and ghost operator have been projected onto these harmonic sectors can one determine which modes are genuine zero modes, slightly-nonzero modes, or ordinary nonzero modes and hence obtain the coefficient of the logarithmic correction.

\section{Soft modes and the one-loop partition function}\label{sec:soft}

We now compute the one-loop correction to $\log Z$ around the
near-extremal saddle at fixed charge $\cQ=\cQ_0$ (equivalently $Q=Q_0$), to leading order in $T$. Only the modes that become exact zero modes at extremality can produce $\log T$, so it suffices to identify these modes and count them. We write the one-loop determinant as
\begin{equation}
\log Z_{1\text{-loop}}=\log Z_{\rm zm}+\log Z_{\rm nz} \ ,
\label{eq:split}
\end{equation}
where $Z_{\rm zm}$ contains the exact zero modes of the extremal throat and $Z_{\rm nz}$ the rest~\cite{Castro:2018hsc,Sen:2012dw,Iliesiu:2022onk}. A mode with extremal eigenvalue
$\lambda_0>0$ has $\log\lambda(T)=\log\lambda_0+\mathcal O(T)$, so
\begin{equation}
\log Z_{\rm nz}=C_{\rm nz}+\mathcal O(T) \ ,
\label{eq:Znz}
\end{equation}
and all $\log T$ dependence comes from $Z_{\rm zm}$. The finite-temperature eigenvalues of these modes can be written schematically as $\lambda_n=\kappa_n T+\mathcal O(T^2)$. The coefficient $\kappa_n$ is a finite matrix element of the fully gauge-fixed finite-temperature Hessian; it depends on the normalization of the fluctuation modes, the boundary conditions and the Lovelock couplings, and zero-mode counting alone does not determine it. We therefore do not assign separate temperature scales to the tensor, vector and $U(1)$ sectors. All such finite constants are absorbed into a single dimensionful reference scale $\mu_T$ and an additive one-loop constant, and, as shown below, they do not affect the coefficient of $\log T$.

In the disc coordinates $\rho=\rho_h\cosh\eta$,
$\theta=2\pi T\tau_E\sim\theta+2\pi$, the finite-temperature throat
Eq.~\eqref{eq:NHnearT} becomes
\begin{equation}
ds^2=L_2^2\big(d\eta^2+\sinh^2\!\eta\,d\theta^2\big)
+\big(r_0+2\pi L_2^2T\cosh\eta\big)^2 d\Omega_5^2+\mathcal O(T)\ ,
\label{eq:disc}
\end{equation}
where the sphere radius follows from $r-r_0=\delta r\cosh\eta$ with
$\delta r=2\pi L_2^2T$. Since $\cosh\eta$ grows without bound, the $\mathcal O(T)$ deformation is small only in the throat region $2\pi L_2^2T\cosh\eta\ll r_0$, i.e.\ $\eta\lesssim\log\!\big(r_0/(L_2^2T)\big)$; beyond it the throat is glued to the exterior geometry, which does not affect the leading zero-mode counting. At $T=0$ the geometry is exactly
$H_2\times S^5$, and the modes identified below are exact zero modes. The $\mathcal O(T)$ deformation in Eq.~\eqref{eq:disc} lifts them, and first-order perturbation theory gives eigenvalues
$\lambda_n=\kappa_n T+\mathcal O(T^2)$, where $n$ is the Fourier label along $\theta$ and $\kappa_n>0$ is temperature independent~\cite{Alvarado:2026kio,Iliesiu:2022onk}. Its $n$-dependence differs between sectors (for example $\kappa_n\propto|n|$ for the bulk gauge towers, and $\kappa_n\propto n^2(n^2-1)$ for the boundary Schwarzian modes), but only the overall power of $T$ matters for the logarithm. Each label $n\neq0$ describes one real mode for each sign of $n$. For a tower with $|n|\ge n_0$ the Gaussian integral therefore gives,
\begin{equation}
\log Z_{\rm tower}=-\frac12\sum_{|n|\ge n_0}\log\big(\kappa_n T\big)
= -\Big(\sum_{n\ge n_0}1\Big)\log T+\text{const} \ ,
\label{eq:towergeneric}
\end{equation}
and with zeta-function regularization
\begin{equation}
\sum_{n\ge1}1=\zeta(0)=-\frac12 \ ,
\qquad
\sum_{n\ge2}1=\zeta(0)-1=-\frac32 \ .
\label{eq:zetasums}
\end{equation}
This is the counting rule used throughout: a tower starting at $n_0=1$ contributes $\tfrac12\log T$, and one starting at $n_0=2$ contributes $\tfrac32\log T$. The coefficients $\kappa_n$ enter only the constant, so we never need them. All the zero modes of Eq.~\eqref{eq:disc} are generated by gauge parameters that are not normalizable on $H_2$, while the resulting fluctuations are. Such modes are not removed by gauge fixing~\cite{Sen:2008vm, Banerjee:2010qc}. They are built from the harmonic functions
\begin{equation}
\Phi_n=e^{in\theta}\Big(\tanh\frac\eta2\Big)^{|n|},
\qquad \nabla^2_{H_2}\Phi_n=0,
\label{eq:Phin}
\end{equation}
whose gradients are square-integrable for $|n|\ge1$. There are three such sectors.

The Schwarzian degree of freedom is a boundary reparametrization $\varphi(\theta)=\theta+\xi(\theta)$ with $\xi(\theta)=\sum_n\xi_n e^{in\theta}$. Its quadratic action is proportional to $\sum_n n^2(n^2-1)|\xi_n|^2$. Quotienting by the $SL(2,\mathbb R)$ redundancy removes $n=0,\pm1$, leaving the tower $|n|\ge2$. This boundary characterization suffices for the counting; a bulk representative must separately satisfy the gauge and normalizability conditions. By Eq.~\eqref{eq:towergeneric} and Eq.~\eqref{eq:zetasums},
\begin{equation}
\log Z_{\rm Sch}=\frac32\log T+\text{const} \ .
\label{eq:ZSch}
\end{equation}
This is the prefactor $(\CL T)^{3/2}$ of Eq.~\eqref{eq:SchZ}. Since the Schwarzian path integral is one-loop exact~\cite{Stanford:2017thb}, Eq.~\eqref{eq:ZSch} holds throughout the throat regime.

For a gauge group $G$ on the near-horizon solution, the large gauge
transformations
\begin{equation}
\delta\mathcal A^I_a=\partial_a\Phi_n=e^{in\theta}\left[\partial_\eta\!\left(\tanh^{|n|}\frac\eta2\right)\delta_a^\eta+in\tanh^{|n|}\frac\eta2\,\delta_a^\theta\right] \ , \quad |n|\ge1,\quad
I=1,\dots,\dim G,
\label{eq:largegauge}
\end{equation}
are soft modes. Two gauge fields are present: the Maxwell field, $G=U(1)$, and the Kaluza–Klein gauge field of the sphere rotations, $G=SO(6)$, defined by $h_{ai}=\mathcal A^{AB}_a\,\mathcal K^{AB}_i$ with $\mathcal K^{AB}_i$ the Killing vectors Eq.~\eqref{Eq:Kiling_vectors}. A Killing vector alone does not deform the sphere; dressed with $\Phi_n$ it produces the normalizable fluctuation $h_{ai}=\partial_a\Phi_n\,\mathcal K^{AB}_i$. Each generator gives a tower with $n_0=1$, so the Gaussian evaluation yields
\begin{equation}
\log Z^{\rm Gauss}_G=\frac{\dim G}{2}\log T+\text{const} \ ,
\label{eq:ZGgauss}
\end{equation}
that is $\tfrac12$ for $U(1)$ and $\tfrac{15}{2}$ for $SO(6)$. Adding the Schwarzian sector, the naive total would be $\tfrac32+\tfrac12+\tfrac{15}{2}=\tfrac{19}{2}$. The next subsection shows that this overcounts in two ways.

\subsection{Exact treatment of the gauge sectors}
\label{sec:exactgauge}

The large gauge modes Eq.~\eqref{eq:largegauge} are the degrees of freedom of a particle moving on the group manifold $G$, with action
$\frac K2\int du\,{\rm Tr}(g^{-1}\partial_ug)^2$~\cite{Iliesiu:2020qvm}. The moment of inertia $K$ sets the energy cost of a black-hole state in representation $R$, namely $C_2(R)/2K$. The exact partition function is
\begin{equation}
Z_G(\beta)=\sum_R(\dim R)^2\exp\Big[-\frac{\beta\,C_2(R)}{2K}\Big],
\label{eq:ZG}
\end{equation}
with limits
\begin{equation}
Z_G\longrightarrow
\begin{cases}
{\rm Vol}(G)\Big(\dfrac{K}{2\pi\beta}\Big)^{\dim G/2}, & T\gg1/K,\\[10pt]
1+\mathcal O\big(e^{-\beta C_{2,\min}/2K}\big), & T\ll1/K,
\end{cases}
\label{eq:ZGlimits}
\end{equation}
where $C_{2,\min}$ is the smallest nonzero Casimir among the representations that appear in the sum, and ${\rm Vol}(G)$ is the volume of the global group $G$ in the metric fixed by the generator normalization. For the $S^5$ rotor this is ${\rm Vol}(SO(6))=\tfrac12{\rm Vol}(SU(4))$, which affects only the temperature-independent constant. The first line reproduces the Gaussian result Eq.~\eqref{eq:ZGgauss}. The second shows that below the scale $1/K$ the sector freezes and gives no power of $T$. The Gaussian count is therefore valid only when the temperature is high enough to excite the sector.

The Maxwell sector at fixed charge. Here the gauge group is $U(1)$; the representations are labeled by the integer charge fluctuation $\mathfrak q\in\mathbb Z$ about $\cQ_0$, measured in units of the charge quantum, with $C_2=\mathfrak q^2$. The role of the moment of inertia is played by the charge susceptibility $K_A=(\partial\cQ/\partial\Phi)_{T=0}$. Fixing the charge projects onto
$\mathfrak q=0$:
\begin{equation}
Z_{U(1)}(\beta,\mathfrak q{=}0)=\int_0^{2\pi}\frac{d\vartheta}{2\pi}
\sum_{\mathfrak q\in\mathbb Z}e^{-\beta \mathfrak q^2/2K_A+i\vartheta \mathfrak q}=1 .
\label{eq:ZU1Q}
\end{equation}
The Maxwell sector thus contributes no $\log T$ at any temperature in the canonical ensemble. At fixed potential it would contribute
$\tfrac12\log T$ for $T\gg1/K_A$.

For the $SO(6)$ sector, we choose an ensemble with fixed electric charge and vanishing angular potentials; angular momentum is summed over rather than fixed. Thus, the full sum Eq.~\eqref{eq:ZG} applies with $\dim SO(6)=15$ and $K=K_J$. The rotor configuration space is $SO(6)$ and its Hilbert space is $L^2(SO(6))$, so the sum includes only those irreducible representations of $SU(4) \cong \mathrm{Spin}(6)$ that descend to $SO(6)$, not every $SU(4)$ representation. By Eq.~\eqref{eq:ZGlimits}, this sector contributes $\tfrac{15}{2}\log T$ for $T\gg1/K_J$ and nothing for $T\ll1/K_J$. Whether the $\tfrac{15}{2}$ is ever visible depends on how $1/K_J$ compares with the other scales, to which we now turn.

\subsection{The crossover temperature and the logarithmic coefficient}
\label{sec:KJ}

The moment of inertia is $K_J=(\partial J/\partial\Omega_H)_{T=0,Q}$, with $\Omega_H$ measured relative to a nonrotating frame at infinity. Define $\lambda=L_{\rm eff}^{-2}$ on the Einstein-connected vacuum branch by
\begin{equation}
 \lambda-a\lambda^2+b\lambda^3=L^{-2},\qquad
 D_\infty=1-2a\lambda+3b\lambda^2>0 \ ,
 \label{eq:vacuumlambda}
\end{equation}
where $a=12\alpha_2$ and $b=24\alpha_3$. Appendix~\ref{app:KK} gives the exact susceptibility to linear order in rotation,
\begin{equation}
 K_J=\frac{\pi^2m_0r_0^2}{8G_7(1+\lambda r_0^2)}
 =c_J\frac{\pi^2r_0^2\Delta}{48G_7}=c_J\,\frac{e_J^{-2}}{r_0},\qquad
 c_J=\frac{6m_0}{\Delta(1+\lambda r_0^2)} \ ,
 \label{eq:KJcJ}
\end{equation}
with $m_0=2r_0^4+\tfrac32 ar_0^2+b+5r_0^6/(2L^2)$. Here $e_J$ is the throat gauge coupling of the $SO(6)$ sector, Eq.~\eqref{eq:eJ}, and $c_J$ is the dimensionless coefficient produced by matching the throat to the exterior geometry. Consequently,
\begin{equation}
 \frac{K_J}{C_L}=\frac{m_0r_0^2}{5\Delta L_2^2(1+\lambda r_0^2)}
 =\frac{2m_0r_0^2W}{5L^2\Delta^2(1+\lambda r_0^2)}.
 \label{eq:KJCL}
\end{equation}
In the asymptotically flat limit,
\begin{equation}
 \left.\frac{K_J}{C_L}\right|_{L\to\infty}
 =\frac{2m_0r_0^2(8r_0^2+3a)}{5\Delta^2}
 \xrightarrow{a,b\to0}\frac{32}{5},\qquad c_J\xrightarrow{a,b\to0}12.
 \label{eq:KJCLflat}
\end{equation}
Thus, in the flat Einstein limit the two scales are parametrically comparable but not numerically identical; there is no parametrically separated window. In the Einstein--AdS limit, with $x=r_0^2/L^2$,
\begin{equation}
 \frac{K_J}{C_L}=\frac{(4+5x)(8+15x)}{5(1+x)}
 \sim15x\quad(x\gg1).
 \label{eq:EinsteinAdSratio}
\end{equation}
For large black holes at fixed $L,a,b$ on a regular AdS branch, one has $m_0\to5r_0^6/(2L^2)$, $W\to15r_0^4$, $\Delta\to r_0^4$ and $1+\lambda r_0^2\to r_0^2/L_{\rm eff}^2$, so the leading result is
\begin{equation}
 \frac{K_J}{C_L}\sim\frac{15\,r_0^2\,L_{\rm eff}^2}{L^4} \ ,
 \label{eq:largeAdSLovelock}
\end{equation}
which reduces to $15r_0^2/L^2$ when $a=b=0$. The throat data depend on the bare radius $L$, while the horizon angular velocity depends on $L_{\rm eff}$. This provides a parametrically wide window $1/K_J\ll T\lesssim1/C_L$, provided all effective-theory restrictions also hold.

The dependence on the cubic coupling follows directly from Eq.~\eqref{eq:KJCL}. In the asymptotically flat case, at fixed $r_0,a$,
\begin{equation}
 \partial_b\log(K_J/C_L)=\frac1{m_0}-\frac6\Delta
 =-\frac{11r_0^4+7ar_0^2+3b}{m_0\Delta}<0
 \quad(a,b\ge0).
 \label{eq:flatmonotonicity}
\end{equation}
Since $b=24\alpha_3$, increasing the cubic coupling narrows the scale separation in this domain. For AdS, at fixed $r_0,L,a$,
\begin{align}
 \partial_b\lambda =-\frac{\lambda^3}{D_\infty}\qquad ;\qquad
 \partial_b\log(K_J/C_L) =\frac1{m_0}-\frac6\Delta
 +\frac{r_0^2\lambda^3}{(1+\lambda r_0^2)D_\infty}.
 \label{eq:AdSmonotonicity}
\end{align}
The last term prevents a universal monotonicity conclusion without specifying the allowed coupling domain.

The soft-sector truncation assumes that no additional perturbation becomes gapless or unstable. Second-order Lovelock equations exclude generic higher-derivative extra fields, but do not by themselves prove stability of every harmonic sector. The conditions $\Delta>0$, $W>0$ and $D_\infty>0$ are necessary regularity conditions for the sectors used here, not a complete stability criterion. The near-extremal expansion also requires $2\pi L_2^2T\ll r_0$, in addition to the low-energy restrictions associated with the throat and omitted modes. Collecting Eq.~\eqref{eq:ZSch}, Eq.~\eqref{eq:ZU1Q} and the $SO(6)$ result,
\begin{equation}
\gamma_{\rm soft}=
\begin{cases}
\dfrac32, & T\ll1/K_J,\\[8pt]
\dfrac32+\dfrac{15}{2}=9, & T\gg1/K_J,
\end{cases}
\label{eq:gammasoft}
\end{equation}
with an additional $+\tfrac12$ at fixed potential for $T\gg1/K_A$. The two values in Eq.~\eqref{eq:gammasoft} are asymptotic limits of a smooth crossover. The leading effective theory partition function, with the extremal energy subtracted, is
\begin{equation}
 Z_Q\simeq\mathcal N e^{S_0}(C_LT)^{3/2}e^{2\pi^2C_LT}\,
 Z_{SO(6)}(K_JT) .
 \label{eq:fullcrossover}
\end{equation}
Here $\mathcal N$ is temperature independent, and $Z_{SO(6)}$ is the sum Eq.~\eqref{eq:ZG} with $K=K_J$, restricted to representations of $SO(6)$. We normalize the rotation generators $L_{AB}$ so that each has integer spectrum, equivalently $E=J^2/2K_J$ for rotation in a single plane; then $C_2=\sum_{A<B}L_{AB}^2$ acts as $-\hat D^2$ on functions on $S^5$. The first rotational excitation is the vector representation $\mathbf{6}$, with $C_{2,\min}=5$ and weight $(\dim R)^2=36$, so $E_{\rm gap}=5/(2K_J)$. Summing over all $SU(4)$ representations would instead admit the spinor $\mathbf{4}$, with $C_2=15/4$, and lower the gap by a factor $4/3$. The global group therefore affects the low-lying spectrum and the detailed crossover, though not the endpoint coefficient $9$. In this sense $1/K_J$ sets the characteristic scale of the crossover, while $E_{\rm gap}$ is its exact first excitation. The rotational entropy throughout the crossover is $(1+T\partial_T)\log Z_{SO(6)}$. Its asymptotic logarithmic coefficients are zero and $15/2$.

The canonical partition function in either asymptotic regime is then
\begin{equation}
\log Z_Q=S_0+2\pi^2C_LT+\gamma_{\rm soft}\log\frac{T}{\mu_T}+C_0+\delta_{\rm rot}+\delta_{\rm EFT},
\label{eq:logZQ}
\end{equation}
where $S_0+2\pi^2C_LT$ is the classical exponent of Eq.~\eqref{eq:SchZ},
whose one-loop prefactor is already included in $\gamma_{\rm soft}$, and $C_0$ collects all temperature-independent one-loop terms. The two remainders are expansions in different dimensionless parameters. $\delta_{\rm EFT}$ denotes omitted corrections to the near-extremal effective theory, whose scaling is set by the throat restrictions above, in particular $L_2^2T/r_0\ll1$. $\delta_{\rm rot}$ is the error in replacing the rotor sum by its asymptotic form. For $K_JT\ll1$ it is exponentially small, $\mathcal O(e^{-C_{2,\min}/2K_JT})$. In the rotational high-temperature regime, $K_JT\gg1$, the short-time heat trace on the fifteen-dimensional compact group gives
\begin{equation}
Z_{SO(6)}(y)=A\,y^{15/2}\Big(1+\frac{\sigma_1}{y}+\frac{\sigma_2}{y^2}+\cdots\Big),\quad
\log Z_{SO(6)}(y)=\log A+\frac{15}{2}\log y+\mathcal O(y^{-1}),\quad y=K_JT,
\label{eq:rotorheat}
\end{equation}
with constants fixed by the group metric and Casimir normalization. The logarithm of the partition function therefore contains the additional asymptotic remainder $\delta_{\rm rot}=\mathcal O\big((K_JT)^{-1}\big)$, which is distinct from corrections to the near-extremal effective theory. The coefficient of $\log T$ remains $3/2+15/2=9$. Equivalently, retaining the exact rotor sum,
\begin{equation}
\log Z_Q=S_0+2\pi^2C_LT+\frac32\log(C_LT)+\log Z_{SO(6)}(K_JT)+\log\mathcal N+\delta_{\rm EFT}.
\label{eq:logZQexact}
\end{equation}
Using $S=(1-\beta\partial_\beta)\log Z$, the linear term doubles while the coefficient of $\log T$ is unchanged, giving
\begin{equation}
S(T,Q_0)=\frac{\pi^3}{4G_7}\big(r_0^5+40\alpha_2r_0^3+360\alpha_3r_0\big)
+4\pi^2C_LT+\gamma_{\rm soft}\log\frac{T}{\mu_T}+C_{1\text{-loop}}+\delta_S \ .
\label{eq:Sfinal}
\end{equation}
Here $C_{1\text{-loop}}=C_0+\gamma_{\rm soft}$, and $\delta_S=(1+T\partial_T)(\delta_{\rm rot}+\delta_{\rm EFT})$: the entropy remainder must be obtained by applying $S=(1+T\partial_T)\log Z$, not assigned directly. In particular, a term $\sigma_1/(K_JT)$ in $\log Z$ cancels from $S$, since $(1+T\partial_T)T^{-1}=0$, so an inverse-temperature correction in $\log Z$ need not survive in the entropy. The limiting logarithmic coefficient is unaffected. The linear term reproduces Eq.~\eqref{eq:CL}. The throat description requires $T\ll1/L_2$, and the logarithm competes with the classical linear term for $T\lesssim1/C_L$. Within the throat the coefficient is $9$ whenever $T\gg1/K_J$, whether or not the logarithm dominates the linear term. A regime with $\gamma_{\rm soft}=9$ in which the logarithm is still comparable to the classical term, $1/K_J\ll T\lesssim1/C_L$, exists only when $K_J\gg C_L$, i.e.\ by Eq.~\eqref{eq:KJCL} for large AdS black holes. For a static horizon $S^{D-2}$ the rotation group is $SO(D-1)$, and the same analysis gives at fixed charge
\begin{equation}
\gamma_D\big|_{T\ll1/K_J}=\frac32,
\qquad
\gamma_D\big|_{T\gg1/K_J}=\frac32+\frac{(D-1)(D-2)}{4}.
\label{eq:gammaD}
\end{equation}
Table~\ref{tab:gamma} lists the cases of interest. Setting $\alpha_3$ or both couplings to zero changes the classical data and the crossover, but not the rotation group, so the entries are the same for Einstein, Gauss--Bonnet and third-order Lovelock gravity.
\begin{table}[t]
\centering
\begin{tabular}{|l|c|c|c|c|}
\hline
Theory & Horizon & $\dim G$ & $\gamma\,(TK_J\ll1)$ & $\gamma\,(TK_J\gg1)$\\
\hline
4D RN & $S^2$ & 3 & $3/2$ & $3$\\
5D EGB & $S^3$ & 6 & $3/2$ & $9/2$\\
7D Lovelock & $S^5$ & 15 & $3/2$ & $9$\\
\hline
\end{tabular}
\caption{Coefficient $\gamma$ of $\log T$ at fixed charge, with $G=SO(D-1)$. In the four- and seven-dimensional cases computed here, a wide window $K_J\gg C_L$ opens only in the large-AdS regime. At fixed potential, each entry acquires $+\tfrac12$ for $T\gg1/K_A$. The rotor lives on the connected group $SO(D-1)$, so only representations that descend from $\mathrm{Spin}(D-1)$ contribute (for example, integer spin for $SO(3)$). This affects the gap but not the limiting coefficients.}
\label{tab:gamma}
\end{table}

For asymptotically flat Reissner--Nordstr\"om, $K_J=C_L$, so the $SO(3)$ and Schwarzian scales coincide. The coefficient is $\tfrac32$ for $T\lesssim1/C_L$, in agreement with \cite{Iliesiu:2020qvm,Iliesiu:2022onk}, and $\tfrac32+\tfrac32=3$ for $1/C_L\ll T\ll1/L_2$, as in Table~\ref{tab:gamma}. The naive count $\tfrac32+\tfrac32+\tfrac12=\tfrac72$ never appears at fixed charge, because the Maxwell sector is projected out by Eq.~\eqref{eq:ZU1Q}. The five-dimensional EGB analysis of Ref.~\cite{Alvarado:2026kio} works at fixed charge but evaluates all towers, including the Maxwell one, in the Gaussian approximation, obtaining $\tfrac32+\tfrac62+\tfrac12=5$. The present treatment differs in the two points of Sec.~\ref{sec:exactgauge}. In the regime $T\gg1/K_J$, where the Gaussian treatment of the $SO(4)$ towers is valid, our counting gives $\tfrac32+3=\tfrac92$ (Table~\ref{tab:gamma}); the difference from Ref.~\cite{Alvarado:2026kio} is then exactly the Maxwell $\tfrac12$ removed by Eq.~\eqref{eq:ZU1Q}. For related discussions of near-extremal logarithms in higher-curvature and higher-dimensional settings see~\cite{Maulik:2024dwq,Maulik:2025hax,Acito:2026mmf,Despontin:2026xzg,Moreno:2022unk,Cheng:2026urz,Kapec:2023ruw,H:2023qko,Gupta:2014hxa,Sen:2012dw,David:2021qaa,Karan_2019,Sen_20122,Bhattacharyya_2012}.

\section{Charged-matter stability and the rotational crossover}
\label{sec:schwinger}

The rotational crossover of Sec.~\ref{sec:KJ} is physically useful only if the charged black-hole background remains sufficiently long-lived. A complementary test is whether its electric field supports Schwinger pair production. The Einstein–Maxwell–Lovelock action by itself contains no electrically charged matter, so this test requires an explicit extension of the field content. We introduce a minimally coupled complex scalar of mass $m_s\geq0$ and charge $g_s$,
\begin{equation}
 I_\Psi=-\int d^7x\sqrt{-g}\,
 \left[(D_\mu\Psi)^*D^\mu\Psi+m_s^2|\Psi|^2\right],
 \qquad D_\mu=\nabla_\mu-i g_s A^{\rm EM}_\mu .
 \label{eq:chargedscalaraction}
\end{equation}
Here $g_s$ is defined relative to the Maxwell potential in Eq.~\eqref{eq:actionfull}; bounds on it must therefore be interpreted in that normalization. We work at quadratic order about $\Psi=0$, neglecting its stress tensor and current in the background equations. The classical quantities $L_2$, $C_L$ and $K_J$ derived above are unchanged at this order. The scalar determinant is an additional quantum contribution, not part of the original graviton–Maxwell mode count. The charged $\AdS_2$ analysis below follows the standard near-horizon treatment of pair production \cite{Chen:2012yn, Chen:2020trh}. 

\subsection{The charged scalar threshold in the throat}

For the Lorentzian near-extremal throat, choose a gauge regular at its outer horizon,
\begin{equation}
 ds_2^2=L_2^2\left[-h(\rho)d\tau^2+\frac{d\rho^2}{h(\rho)}\right],
 \qquad h(\rho)=\rho^2-\rho_h^2,
 \qquad A^{\rm EM}=e(\rho-\rho_h)d\tau,
 \label{eq:schwingerthroat}
\end{equation}
where $e=q_0L_2^2/r_0^5$ is signed, so that $F=e\,d\rho\wedge d\tau$. This wedge ordering follows from $F_{rt}=q_0/r^5$ in the static convention. The physical electric-field magnitude is
\begin{equation}
 E_0=\frac{|q_0|}{r_0^5}=\frac{\sqrt{10}|Q_0|}{r_0^5},
 \qquad |e|=E_0L_2^2.
 \label{eq:schwingerelectric}
\end{equation}
Separating $\Psi=e^{-i\varpi\tau}R_\ell(\rho)Y^{(\ell,\mathcal A)}$ gives
\begin{equation}
 \frac{d}{d\rho}\left(h\frac{dR_\ell}{d\rho}\right)
 +\left[\frac{\left(\varpi+g_s e(\rho-\rho_h)\right)^2}{h}
 -M_\ell^2L_2^2\right]R_\ell=0,
 \qquad M_\ell^2=m_s^2+\frac{\ell(\ell+4)}{r_0^2}.
 \label{eq:schwingerradial}
\end{equation}
At the formal $\AdS_2$ boundary the two radial behaviors are $R_\ell\sim\rho^{-1/2\pm\nu_\ell}$, with
\begin{equation}
 \nu_\ell^2=\frac14+M_\ell^2L_2^2-g_s^2E_0^2L_2^4.
 \label{eq:schwingerindex}
\end{equation}
Consequently $\nu_\ell^2<0$ violates the charged $\AdS_2$ Breitenlohner--Freedman (BF) bound and opens the oscillatory, flux-carrying channel used to describe throat Schwinger production under pair-production boundary conditions. The surface $\nu_\ell^2=0$ is its threshold. For $\nu_\ell^2>0$ that BF-violating channel is absent. The index is independent of $\rho_h$ at leading throat order, whereas finite-temperature production rates depend on the horizon boundary condition and on frequency. This criterion therefore does not state that all charged emission vanishes at finite temperature.

Since $\nu_\ell^2-\nu_0^2=\ell(\ell+4)L_2^2/r_0^2\geq0$, the $s$ wave is the first mode to violate the bound. For this scalar species the boundary is
\begin{equation}
 g_s^2E_0^2L_2^4=m_s^2L_2^2+\frac14.
 \label{eq:schwingerthreshold}
\end{equation}
Absence of an $s$-wave violation implies absence of this violation for all its higher spherical harmonics. It does not prove stability of the gravitational sectors or of other matter species. Nor does a local throat criterion determine the flux reaching the asymptotic region: that requires matching through the exterior and specifying the asymptotic boundary conditions.

\subsection{An analytic boundary in the cubic coupling}

For the spherical extremal solution, Eq.~\eqref{eq:Q0} gives
\begin{equation}
 E_0^2=\frac{10}{r_0^2}+\frac{5a}{r_0^4}+\frac{15}{L^2},
 \qquad
 L_2^2=\frac{\Delta}{2H},\quad
 H=8r_0^2+3a+\frac{15r_0^4}{L^2},\quad
 \Delta=r_0^4+2ar_0^2+3b.
 \label{eq:schwingerlovelockdata}
\end{equation}
Unlike $L_2$, the electric field is independent of $b$ at fixed $r_0,a,L$. Assume $g_sE_0\neq0$ and $H>0$. The positive solution of Eq.~\eqref{eq:schwingerthreshold}, viewed as a quadratic equation in $z=L_2^2$, is
\begin{equation}
 z_c=\frac{m_s^2+\sqrt{m_s^4+g_s^2E_0^2}}{2g_s^2E_0^2},
 \qquad
 b_{\rm crit}=\frac{2Hz_c-r_0^4-2ar_0^2}{3}.
 \label{eq:schwingerbcrit}
\end{equation}
On the portion of a regular branch with $\Delta>0$, and at fixed $r_0,a,L,m_s,g_s$, the $s$ wave has $\nu_0^2>0$ for $b<b_{\rm crit}$ and $\nu_0^2<0$ for $b>b_{\rm crit}$. In the original action normalization, $\alpha_{3,\rm crit}=b_{\rm crit}/24$. This is a boundary only if it lies inside the admissible coupling domain: it must be intersected with the vacuum and exterior regularity conditions and with any independent gravitational stability or effective-theory restrictions. If it lies outside that domain, no accessible transition follows. When $g_sE_0=0$, the minimally coupled scalar with $m_s^2\geq0$ has no such charged BF violation.

The cubic interaction therefore supplies a direct connection between two effects studied here. Increasing $b$ at fixed $r_0,a,L$ increases $L_2^2$ and eventually reaches the scalar threshold if it is accessible. In the asymptotically flat nonnegative-coupling domain, Eq.~\eqref{eq:flatmonotonicity} simultaneously shows that $K_J/C_L$ decreases. Along this family, increasing the cubic coupling both reduces the rotational scale separation and can open a charged-pair channel. No corresponding universal monotonicity statement for $K_J/C_L$ is implied in AdS, because of the additional vacuum-radius term in Eq.~\eqref{eq:AdSmonotonicity}.

\subsection{Compatibility with the large-AdS rotational window}

The parametrically broad rotational window of Sec.~\ref{sec:KJ} occurs for large AdS black holes. At fixed $L,a,b$ and $r_0/L\to\infty$,
\begin{equation}
 L_2^2\longrightarrow\frac{L^2}{30},\qquad
 E_0^2\longrightarrow\frac{15}{L^2},\qquad
 \frac{K_J}{C_L}\sim\frac{15r_0^2L_{\rm eff}^2}{L^4}\gg1.
 \label{eq:schwingerlargeads}
\end{equation}
For fixed particle parameters the $s$-wave index becomes
\begin{equation}
 \nu_0^2=\frac14+\frac{L^2}{60}(2m_s^2-g_s^2)+o(1),
 \qquad r_0/L\to\infty.
 \label{eq:schwingerlargeindex}
\end{equation}
Thus the limiting boundary is
\begin{equation}
 g_s^2=2m_s^2+\frac{15}{L^2}.
 \label{eq:schwingerlargeboundary}
\end{equation}
Below this value the added scalar has no BF-violating throat channel, whereas above it such a channel is present for sufficiently large $r_0$. Exactly on the limiting surface, subleading finite-radius terms determine the sign. The leading boundary has no explicit Lovelock-coupling dependence, even though the rotational hierarchy retains $L_{\rm eff}$ dependence. Hence a wide rotational regime does not by itself imply a Schwinger instability: the two conditions test different aspects of the same throat.

The relevant intersection for the entropy crossover is therefore
\begin{equation}
 K_J^{-1}\ll T\lesssim C_L^{-1},\qquad
 T\ll T_{\rm EFT},\qquad
 2\pi L_2^2T\ll r_0,\qquad
 \nu_0^2>0,
 \label{eq:schwingerrotorintersection}
\end{equation}
where $T_{\rm EFT}$ denotes the cutoff set by omitted modes and higher-order corrections, and the scalar is assumed to remain separated from its threshold. These conditions identify a candidate region for the $\gamma_{\rm soft}=9$ result, subject to the gravitational stability assumptions already stated. They are not a full phase diagram. With several charged species, the corresponding absence-of-violation condition must hold for each species; fermions and nonminimally coupled fields require their own fluctuation analysis.

Finally, fixed total electric charge does not forbid pair production: a pair has zero net charge, while its separation can transfer charge between the horizon and the exterior. Thus a canonical ensemble at infinity does not fix the horizon charge in the presence of a charged cloud. If production is allowed but strongly suppressed, the original saddle may still describe a metastable black hole. An operational condition is $\tau_{\rm rel}\ll\tau_Q$, with $\tau_Q=|Q_H/\dot Q_H|$ the horizon-charge evolution time and $\tau_{\rm rel}$ the relaxation time of the retained sectors. In particular, $T\tau_Q\gg1$ is needed for negligible discharge during a thermal time. Computing $\tau_Q$ requires the production rate, exterior matching and backreaction; it is not determined by Eq.~\eqref{eq:schwingerindex}. Close to threshold, or when discharge is rapid, the charged determinant and its coupling to the soft sectors must be retained explicitly. The original coefficients $3/2$ and $9$ cannot then be asserted to describe the complete enlarged theory without that analysis.

\section{Discussion and conclusions}
\label{sec:conclusions}

We have computed the low-temperature one-loop correction to the entropy of a static charged black hole in seven-dimensional Einstein–Maxwell gravity with quadratic and cubic Lovelock terms. Seven is the lowest dimension in which the cubic Lovelock density is dynamical, so this is the simplest setting in which to ask how a genuine six-derivative curvature interaction enters the near-extremal problem. We find that the Lovelock couplings change the classical saddle and the temperature at which each soft sector becomes active, while leaving the number of soft modes unchanged.

At the classical level, the extremal near-horizon geometry is $\AdS_2\times S^5$, with the radius $L_2$ given exactly by Eq.~\eqref{eq:L2sphere}. Near extremality $r_+-r_0=2\pi L_2^2T+\mathcal O(T^2)$, and the classical entropy is $S_0+4\pi^2\CL T+\mathcal O(T^2)$. The cubic coupling enters all of these quantities: it appears as $360\be r_0$ in the Wald entropy, as $24\be$ in the extremal mass, and through $\Delta$ in $L_2$ and $\CL$. The one-loop determinant is therefore evaluated on a geometry that depends explicitly on both Lovelock couplings.

At one loop, the infrared logarithm comes from three soft sectors. Each is a tower of modes that are exact zero modes of the extremal throat and are lifted linearly in $T$. The Schwarzian tower of boundary reparametrizations starts at $|n|=2$ and contributes $\tfrac32$ at all temperatures. This is the same $\tfrac32$ that appears in the Schwarzian prefactor, and it does not depend on the value of $\CL$. The Maxwell tower is projected out by the charge constraint Eq.~\eqref{eq:ZU1Q} and contributes nothing in the canonical ensemble. The fifteen towers associated with the Killing vectors of $S^5$ form a particle on $SO(6)$, which by Eq.~\eqref{eq:ZGlimits} contributes $\tfrac{15}{2}$ only when the temperature exceeds the scale $1/K_J$. The logarithmic part of Eq.~\eqref{eq:Sfinal} is therefore
\begin{equation}
S_{\log}=
\begin{cases}
\tfrac32\log(T/\mu_T)+C_{1\text{-loop}}+\delta_S, & T\ll1/K_J,\\[4pt]
9\log(T/\mu_T)+C_{1\text{-loop}}+\delta_S, & T\gg1/K_J .
\end{cases}
\label{eq:DeltaSlog}
\end{equation}
Here $\delta_S$ is the remainder defined below Eq.~\eqref{eq:Sfinal}, which keeps the near-extremal and rotor errors separate. At fixed potential both values acquire an additional $\tfrac12$ above the Maxwell gap. At fixed charge, the naive Gaussian count $\tfrac{19}{2}$ is never the correct coefficient, because it includes the Maxwell mode, which is projected out at fixed charge, and ignores the $SO(6)$ gap.

The Lovelock couplings do not enter these multiplicities. The values $\tfrac32$ and $\tfrac{15}{2}$ are fixed by the symmetries of the near-horizon geometry, namely the asymptotic symmetries of $\AdS_2$ and the isometry algebra $\mathfrak{so}(6)$ of the horizon. Lovelock gravity has second-order field equations, so on a regular branch its spectrum is still one graviton and one Maxwell field. There is no additional scalar or massive spin-2 mode that could supply new soft sectors. The couplings enter instead through the radial kinetic coefficient and the susceptibility, which we determine analytically,
\begin{equation*}
 K_J=\frac{\pi^2m_0r_0^2}{8G_7(1+\lambda r_0^2)},\qquad
 \frac{K_J}{C_L}=\frac{2m_0r_0^2W}{5L^2\Delta^2(1+\lambda r_0^2)} \ .
\end{equation*}
In the flat Einstein limit $K_J/C_L=32/5$, so the two scales are parametrically comparable, though not equal. For large Einstein--AdS black holes $K_J/C_L\sim15r_0^2/L^2$, and with Lovelock couplings $K_J/C_L\sim15r_0^2L_{\rm eff}^2/L^4$, so the rotational window is parametrically separated. More generally, the hierarchy is set by Eq.~\eqref{eq:KJCL}, subject to regularity and stability. At fixed $r_0$ and $a$, increasing $b=24\alpha_3$ lowers this ratio for asymptotically flat black holes with nonnegative couplings. In AdS the vacuum-radius contribution in Eq.~\eqref{eq:AdSmonotonicity} must also be included. The two logarithmic coefficients are the limits of the smooth crossover Eq.~\eqref{eq:fullcrossover}; the coefficient does not jump between them.

As a consistency check, the same reasoning in four dimensions gives $K_J=\CL$ exactly for asymptotically flat Reissner--Nordstr\"om (Appendix~\ref{app:4D}). This reproduces the established fixed-charge coefficient $\tfrac32$ at $T\lesssim1/\CL$, which any counting scheme must recover. In five-dimensional EGB gravity the horizon $S^3$ has six Killing vectors, so the coefficient in the corresponding window is $\tfrac92$. The seven-dimensional value exceeds it by $\tfrac92$, one half for each of the nine additional Killing vectors of $S^5$. Below their respective gaps, both theories give $\tfrac32$.

The coefficient of $\log T$ is an infrared quantity, determined by eigenvalues that vanish at extremality. It should be kept separate from logarithms of geometric scales such as $r_0$, which come from the local heat-kernel expansion of the gapped spectrum~\cite{Modak:2025gvp}. In odd bulk dimension the bulk logarithmic Seeley--DeWitt coefficient vanishes~\cite{Vassilevich:2003xt,Sen:2012dw,David:2009xg}, so the soft logarithm cannot be identified with a local seven-dimensional coefficient. The coefficient is universal only within the stated ensemble, temperature window and field content. The result fails at degenerate Lovelock points, where $\Delta\to0$, the graviton kinetic term vanishes and additional modes can become light. It also fails near $W\to0$, where the $\AdS_2$ expansion breaks down, and it can change if further matter, scalar moduli, additional gauge fields or a supersymmetric completion are included. The finite constant $C_{1\text{-loop}}$ depends on the full nonzero-mode determinant and on normalization conventions, and the present analysis does not determine it.

The probe charged-scalar analysis of Sec.~\ref{sec:schwinger} gives an independent check on the rotational crossover. We derive the threshold $g_s^2E_0^2L_2^4=m_s^2L_2^2+1/4$ and the corresponding critical cubic coupling, Eq.~\eqref{eq:schwingerbcrit}. For large AdS black holes the boundary reduces to $g_s^2=2m_s^2+15/L^2$, independent of the Lovelock couplings at leading order. Below this boundary, the broad rotational window can coexist with the absence of this scalar's BF-violating throat channel. The test covers only this one sector. A full stability analysis, and the discharge timescale when pair production is allowed, are still needed before the entropy crossover can be read as an equilibrium effect.

Several questions are left open. For the grand-canonical ensemble one needs the charge susceptibility $K_A=(\partial \cQ/\partial\Phi)_{T=0}$. Stability of the full harmonic spectrum remains to be established, and the finite-temperature Hessian would fix the signs and normalizations of the lifted eigenvalues. A controlled derivation of corrections to the leading Schwarzian-plus-rotor factorization would show how far it can be trusted. Hyperbolic horizons and additional matter are further settings where the soft-sector content would have to be worked out again.

\section*{Acknowledgements}
Ankit Anand is financially supported by the Institute's postdoctoral fellowship at IITK and by the National Postdoctoral Fellowship (NPDF) of the Anusandhan National Research Foundation (ANRF), India, under Grant No. PDF/2026/003447, IIT Bhubaneswar. Aditya Singh is grateful to CSIR-HRDG for the financial support received under Project No. 03WS(003)/2023-24/EMR-II/ASPIRE. The authors used Claude (Anthropic) for language editing and as an aid in cross-checking algebraic steps. All results were verified independently by the authors, who take full responsibility for the content of this paper.

\appendix

\section{Four-dimensional moment of inertia}\label{app:4D}

We check the structure of Eq.~\eqref{eq:KJCL} in four dimensions, where the moment of inertia can be computed exactly. Consider
Reissner--Nordstr\"om--AdS with $G=1$,
\begin{equation}
f(r)=1-\frac{2M}{r}+\frac{Q^2}{r^2}+\frac{r^2}{L^2},
\qquad x=\frac{r_0^2}{L^2} \ .
\end{equation}
The extremality conditions $f(r_0)=f'(r_0)=0$ give $Q_0^2=r_0^2(1+3x)$, $M_0=r_0(1+2x)$ and then $f''(r_0)=2(1+6x)/r_0^2$, so that
\begin{equation}
L_2^2=\frac{r_0^2}{1+6x},\qquad
C_L=\frac{L_2^2}{2\pi}\,S'(r_0)=r_0L_2^2 \ ,
\end{equation}
using $S=\pi r_+^2$. Slowly rotating Kerr–Newman–AdS has, to first order in the rotation parameter $a$ \cite{Caldarelli:1999xj},
\begin{equation}
J=aM_0,\qquad \Omega_H=\frac{a(1+x)}{r_0^2},
\end{equation}
so the moment of inertia at fixed charge is
\begin{equation}
K_J=\frac{J}{\Omega_H}=r_0^3\,\frac{1+2x}{1+x},
\qquad
\frac{K_J}{C_L}=\frac{1+2x}{1+x}\,\frac{r_0^2}{L_2^2}
=\frac{(1+2x)(1+6x)}{1+x}.
\label{eq:4Dcheck}
\end{equation}
This has the form Eq.~\eqref{eq:KJCL}, a factor $r_0^2/L_2^2$ times a coefficient $c(x)=(1+2x)/(1+x)$ bounded between $1$ and $2$. In the flat limit $x\to0$ it gives $K_J=C_L=Q_0^3$ \cite{Iliesiu:2020qvm}, and
for large black holes $K_J/C_L\simeq12\,r_0^2/L^2$.

\section{Slow rotation and the rotational susceptibility}
\label{app:KK}

We determine the rotational susceptibility by matching the coupled gravitational and Maxwell perturbations through the exterior geometry. We use the charged slow-rotation solution of~\cite{Yue:2011et} (see also~\cite{Kim:2007iw} for the Gauss--Bonnet case), translated to the normalization of Eq.~\eqref{eq:actionfull}.
The rotation parameter is denoted by $\omega$, distinct from $a=12\alpha_2$. Introduce $\psi=(1-f)/r^2$ and $\Upsilon(\psi)=\psi+a\psi^2+b\psi^3$. The static equation is
\begin{equation}
 \Upsilon(\psi)=-L^{-2}+mr^{-6}-Q^2r^{-10},\qquad q^2=10Q^2 \ .
 \label{eq:rotationpolynomial}
\end{equation}
For one rotation plane, the fibred metric is
\begin{equation}
 ds^2=g_{ab}\dd x^a\dd x^b+r^2[\dd\theta^2+\cos^2\theta(\dd\phi+A)^2+\sin^2\theta\dd\Omega_3^2] \ ,
 \label{eq:KKansatz}
\end{equation}
with $A=A_t(r)\dd t$ and $g_{t\phi}=r^2\cos^2\theta A_t$. The elementary rotation-plane generators are normalized equally by spherical symmetry. The Maxwell potential is
\begin{equation}
 A^{\rm EM}=-\frac{q}{4r^4}\dd t+\chi(r)\cos^2\theta\dd\phi,
\end{equation}
up to a constant electric-potential shift. The radial coefficient in the linearized gravitational equation is
\begin{equation}
 \cP(r)=r^7\Upsilon'(\psi)
 =r^7+24\al r^5(1-f)+72\be r^3(1-f)^2.
 \label{eq:Pfull}
\end{equation}
This expression concerns the radial slow-rotation problem; it is not a derivation of the complete off-shell Lovelock Hessian. Its horizon value is
\begin{equation}
 \cP(r_0)=r_0^3\Delta,\qquad
 S_{\rm W}'(r_0)=\frac{5\pi^3\Delta}{4G_7}.
 \label{eq:Pnh}
\end{equation}
Using the Killing-vector normalization in Eq.~\eqref{eq:Knorm}, the corresponding throat gauge kinetic term is written as
\begin{equation}
 I^{(2)}_{\rm vec}=-\frac1{4e_J^2}\int\dd^2x\sqrt{-g_2}\,F_{ab}F^{ab} \quad \text{with}\quad e_J^{-2}=\frac{\pi^2\cP(r_0)}{48G_7} =\frac{\pi^2r_0^3\Delta}{48G_7} =\frac{r_0^3}{60\pi}S_{\rm W}'(r_0) \ .
\label{eq:eJ}
\end{equation}

The moment of inertia of the $SO(6)$ rotor is $K_J=c_J\,e_J^{-2}/r_0$, Eq.~\eqref{eq:KJcJ}, where the coefficient $c_J$ is fixed by the exterior solution derived below. Up to the common normalization and radial boundary terms, a compatible reduced Lagrangian is
\begin{equation}
 \mathcal L=\tfrac12\cP A_t'^2-4qA_t\chi'-2fr^3\chi'^2-16r\chi^2.
 \label{eq:slowrot}
\end{equation}
The mixing term is equivalent to $+4qA_t'\chi$ after integration by parts. Variation gives
\begin{equation}
 \cP A_t'+4q\chi=C,\qquad
 (fr^3\chi')'=-qA_t'+8r\chi.
 \label{eq:chieq}
\end{equation}
The first equation is the integrated gravitational equation. With $\lambda$ defined in Eq.~\eqref{eq:vacuumlambda}, an explicit solution is $ A_t=-\omega(\psi+\lambda)$ and $ \chi=\omega q/4r^4.$ It obeys $A_t(\infty)=0$ and has a decaying magnetic perturbation. To check the gravitational equation, differentiate Eq.~\eqref{eq:rotationpolynomial}:
\begin{equation}
 \Upsilon'(\psi)\psi'=-\frac{6m}{r^7}+\frac{10Q^2}{r^{11}}.
\end{equation}
Consequently,
\begin{equation}
 \cP A_t'=\omega\left(6m-\frac{10Q^2}{r^4}\right),\qquad
 4q\chi=\frac{10\omega Q^2}{r^4},\qquad C=6\omega m.
\end{equation}
For the Maxwell equation, both sides agree explicitly:
\begin{align}
 (fr^3\chi')'=-\omega q\left(\frac{f'}{r^2}-\frac{2f}{r^3}\right)\quad \text{and}\quad
 -qA_t'+8r\chi =\omega q\left(\psi'+\frac2{r^3}\right)
 =-\omega q\left(\frac{f'}{r^2}-\frac{2f}{r^3}\right).
\end{align}
The angular momentum and horizon angular velocity relative to infinity are
\begin{equation}
 J=\frac{\pi^2C}{48G_7}=\frac{\pi^2\omega m}{8G_7},\qquad
 \Omega_H=-A_t(r_+)=\omega(r_+^{-2}+\lambda),\qquad
 K_J=\left(\frac{\partial J}{\partial\Omega_H}\right)_{T=0,Q}.
 \label{eq:KJdef}
\end{equation}
Along the fixed-charge extremal family, $r_+$ and $m$ change only at second order in $\omega$. Thus their static extremal values suffice to determine the linear susceptibility:
\begin{equation}
 K_J=\frac{\pi^2m_0r_0^2}{8G_7(1+\lambda r_0^2)}.
 \label{eq:KJanalytic}
\end{equation}
The integral representation
\begin{equation}
 A_t(r)=-\int_r^\infty\dd r'\frac{C-4q\chi(r')}{\cP(r')}
 \label{eq:Atint}
\end{equation}
can then be evaluated analytically as
\begin{equation}
 \int_{r_0}^\infty\frac{\dd r}{\cP(r)}\left(1-\frac{4q\chi}{C}\right)
 =\frac{r_0^{-2}+\lambda}{6m_0},\qquad
 c_J=\frac{6m_0}{\Delta(1+\lambda r_0^2)} \ .
 \label{eq:cJ}
\end{equation}
This determines the exterior matching coefficient $c_J$ and leads directly to the crossover ratio in Eq.~\eqref{eq:KJCL}.

\bibliographystyle{JHEP}
\bibliography{ref_lov}

\end{document}